\documentclass[conference]{IEEEtran}

\usepackage{amsmath,amssymb}
\usepackage{booktabs}
\usepackage{graphicx}
\usepackage{hyperref}
\usepackage{cleveref}
\usepackage{algorithm}
\usepackage{algpseudocode}
\usepackage{xcolor}
\usepackage{enumitem}
\usepackage{caption}
\usepackage{listings}
\usepackage{microtype}
\usepackage{multirow}
\usepackage{tabularx}
\usepackage{pifont}
\usepackage{tikz}

\hypersetup{
  colorlinks=true,
  linkcolor=blue!70!black,
  citecolor=green!50!black,
  urlcolor=blue!60!black,
}

\crefname{lstlisting}{Listing}{Listings}
\Crefname{lstlisting}{Listing}{Listings}

\newcommand{\ncsim}{\texttt{ncsim}}

\newcommand{\cmark}{\ding{51}}
\newcommand{\xmark}{\ding{55}}

\title{\ncsim{}: A Lightweight Simulator for Networked\\Edge Computing with Wireless Interference Modeling}
\author{
\IEEEauthorblockN{Bhaskar Krishnamachari\IEEEauthorrefmark{1},
Maya Gutierrez\IEEEauthorrefmark{1}, and
Jared Coleman\IEEEauthorrefmark{1}\IEEEauthorrefmark{2}}
\IEEEauthorblockA{\IEEEauthorrefmark{1}University of Southern California\\
bkrishna@usc.edu, mayaguti@usc.edu}
\IEEEauthorblockA{\IEEEauthorrefmark{2}Loyola Marymount University\\
jared.coleman@lmu.edu}
}

\begin{document}
\maketitle

\begin{abstract}
Evaluating directed acyclic graph (DAG) task schedulers for wireless edge computing requires
jointly modeling compute placement and wireless interference, yet
many evaluation setups simplify the wireless coupling. This can lead to
\emph{rank inversions}: a scheduler selected under an interference-free
model need not remain best when concurrent wireless traffic is included.
We present \ncsim{}, a lightweight discrete-event simulator that combines
DAG workflow scheduling with an analytical IEEE~802.11 contention and
hidden-interference model in a single Python package. A 108-run factorial
experiment on grid networks and DAG workflows finds rank inversions
in 6 of 18 scenarios (33.3\%). Solo retains single-link MAC overhead,
while Full adds concurrent-link wireless effects. Selecting a policy under
Solo and executing it under Full incurs a maximum scheduling regret
(makespan penalty relative to the best evaluated policy under Full) of
52.3\%. Validation includes Bianchi's published numerical and
graphical results and fixed-rate ns-3 experiments.
The default model closely tracks homogeneous contention.
The scheduler results show how wireless coupling can change policy
selection within the stated flow model.

\end{abstract}

\section{Introduction}
\label{sec:intro}

Edge and fog computing architectures distribute computation across
heterogeneous nodes connected by networks of varying capacity and
reliability~\cite{shi2016edge}.  Effective task scheduling in these
environments requires co-optimizing compute placement and network transport:
a task placement that minimizes computation time may incur excessive data
transfer costs, while a network-aware placement may underutilize fast
processors.

Existing simulation tools address parts of this problem. Network simulators
such as ns-3~\cite{ns3} and OMNeT++ with its network model
libraries~\cite{omnetpp} provide detailed protocol stacks, while
SimGrid~\cite{simgrid}, WRENCH~\cite{wrench}, and CloudSim~\cite{cloudsim}
model distributed applications and compute resources. SimGrid also models
bandwidth sharing among concurrent flows and within WiFi cells, and offers
an ns-3 backend~\cite{simgridmodels,flowwifi2022}. Edge-specific
tools like iFogSim~\cite{ifogsim} and EdgeCloudSim~\cite{edgecloudsim} add
domain-specific placement and network models. The combination we study is
DAG placement with multihop routes, overlapping carrier-sense neighborhoods,
and service rates affected by hidden transmitters in a Python flow simulator.

We present \ncsim{},\footnote{\url{https://github.com/ANRGUSC/ncsim}}
an open-source, lightweight, flow-level discrete-event simulator that
combines DAG workflow scheduling with analytical 802.11 WiFi
models in a single Python package.  \ncsim{} occupies the
space between packet-level network simulators (ns-3, OMNeT++) and
application and resource simulators (SimGrid, WRENCH, CloudSim, iFogSim): a configurable environment for comparing scheduler outcomes under
explicit wireless assumptions.

To examine the effect on scheduler selection, we use \ncsim{} in a
108-run factorial study comparing Solo, which retains single-link MAC
overhead, with Full, which adds concurrent-link wireless effects.
The scheduler sets that minimize makespan under the two wireless modes
are disjoint in 6 of 18 scenarios. Selecting a policy under Solo incurs a
maximum makespan penalty of 52.3\% under Full, on the $4\times4$ grid with five tasks
and widest-path routing. This comparison isolates concurrent wireless
effects from single-link MAC overhead and shows how they change scheduler
selection in the flow model.

The contributions are:

\begin{enumerate}[itemsep=2pt,leftmargin=*]
\item \textbf{\ncsim{} simulator:} a modular architecture combining DAG
  scheduling through SAGA and built-in policies with log-distance path loss,
  table-based rates, Bianchi contention, service rates affected by hidden
  transmitters, and multihop routing with dynamic bandwidth sharing.
\item \textbf{Three validation stages:} internal component and
  event-engine checks, reproduction of Bianchi's published
  results~\cite{bianchi2000}, and ns-3 comparisons with measurement
  settings and approximation-specific errors stated separately.
\item \textbf{Rank inversion finding:} a 108-run grid
  factorial yields six rank inversions and a mean scheduling regret of 5.6\%.
\item \textbf{Sensitivity:} a CCR sweep and concurrent-workflow
  experiment characterize the effects of payload size and workflow count.
\end{enumerate}

\Cref{sec:related,sec:arch,sec:model,sec:engine} survey related work and
describe \ncsim{}'s architecture, system model, and simulation engine.
\Cref{sec:validation} validates the WiFi model, and \Cref{sec:eval}
presents the experimental evaluation. \Cref{sec:conclusion} concludes.

\section{Related Work}
\label{sec:related}

We situate \ncsim{} at the intersection of DAG scheduling research and
wireless network simulation. Our target users need a combination of
DAG-aware scheduling, wireless CSMA/CA and hidden-terminal modeling,
and lightweight parameter sweeps in one
tool.  \Cref{tab:sim_comparison} summarizes the comparison.

\textbf{DAG scheduling and workflow systems.}
DAG scheduling has a rich history in parallel computing, from
list-scheduling and dynamic-level scheduling to HEFT and
CPOP~\cite{Adam1974ListSchedules,KwokAhmad1999Survey,Sih1993DLS,heft}.
Benchmarking and parametric scheduler design tools such as SAGA, PISA,
and parametric decomposition enable adversarial search and systematic
ablation~\cite{Coleman2024PISA,SAGARepo,Coleman2024Parametric}.
Dispersed and edge execution systems (Jupiter, Tactical Jupiter, and
throughput-oriented extensions) integrate DAG scheduling with
profiling and orchestration~\cite{Ghosh2021Jupiter,Poylisher2021TacticalJupiter,Zhao2021TPHEFT}.
Tactical Jupiter accounts for CSMA interference in its bandwidth estimates
and evaluates an 802.11n MANET using hybrid network
emulation~\cite{Poylisher2021TacticalJupiter}. Our focus is an analytical
flow simulator that couples DAG execution with wireless service rates.

\textbf{Network simulators and wireless/edge DAG systems.}
ns-3~\cite{ns3} and OMNeT++ with network model libraries such as
INET~\cite{omnetpp} provide WiFi and TCP/IP models, but they are
packet-level tools: integrating the placement and execution semantics of
a DAG requires a workflow layer. Their event counts depend on packet
traffic, whereas \ncsim{} schedules changes in flow service. 

A growing body of edge/IoT scheduling and offloading
work tackles heterogeneous devices and mobile edge computing (MEC)
offloading~\cite{Suryavansh2020IBOT,Li2022IBDASH,Li2024MTEC,Yi2017LAVEA,Zhang2019HeteroEdge,Lin2019Petrel,Liang2021JointOffloading,Li2024EnergyConstrained,Long2025SecDS}.
For example, MTEC, LAVEA, and Hetero-Edge use measured network conditions
to estimate communication costs~\cite{Li2024MTEC,Yi2017LAVEA,Zhang2019HeteroEdge}.
Our flow model explicitly represents CSMA/CA and hidden-terminal effects
in 802.11 channels. Related work also
examines broader fog-computing opportunities~\cite{Chiang2016FogIoT,Chiang2017Fog10Q}
and coded edge computing~\cite{Kim2020CodedEdge}.

\textbf{Workflow and edge simulators.}
SimGrid~\cite{simgrid} simulates computation and communication in distributed
applications. Its flow models allocate rates subject to shared-resource
capacity constraints and update ongoing transfers when competing activities
change~\cite{simgridmodels}. The native WiFi model of
Courageux-Sudan et al.~\cite{flowwifi2022} represents an access point and
its stations as a shared wireless channel. It accounts for heterogeneous
station rates and reduces channel capacity as concurrency increases, using
parameters calibrated against ns-3 for the chosen MCS and RTS/CTS settings.
The study evaluates throughput agreement and large-scale simulation;
its modeled scope excludes hidden nodes and inter-cell interference.

\ncsim{} addresses multihop wireless DAG execution with overlapping
carrier-sense and hidden-interference neighborhoods. Node positions and RF
parameters determine link rates and which links interact; local Bianchi
calculations and received interference power determine service as transfers
start and finish. This lets a scheduler's placement and routing choices
change the wireless load experienced by other DAG dependencies within one
Python simulation. Both tools use flow-level updates, but their native
wireless models represent different interactions. SimGrid also offers an
ns-3 backend for packet-level networking~\cite{simgridmodels}.
WRENCH~\cite{wrench} supplies workflow-management abstractions over SimGrid,
while CloudSim~\cite{cloudsim} models cloud applications and resources.

Edge-specific tools such as iFogSim~\cite{ifogsim} and
EdgeCloudSim~\cite{edgecloudsim} add domain-specific workload and placement
features. Our companion MILCOM study~\cite{milcom2026} uses \ncsim{} to study
locality-aware scheduling and routing.

\begin{table}[t]
\centering
\caption{Simulator comparison. WiFi and contention entries describe the
available modeling abstraction; the Packets column identifies packet-level simulation.
DAG indicates a bundled workflow interface; integrations can extend it.
OMNeT++ entries refer to network model libraries; SimGrid entries refer
to native flow models, excluding its ns-3 backend.}

\label{tab:sim_comparison}
\setlength{\tabcolsep}{3.5pt}
\begin{tabular}{@{}lccccc@{}}
\toprule
\textbf{Tool} & \rotatebox{70}{\textbf{DAG}} &
\rotatebox{70}{\textbf{WiFi}} & \rotatebox{70}{\textbf{Contention}} &
\rotatebox{70}{\textbf{Hidden intf.}} & \rotatebox{70}{\textbf{Packets}} \\
\midrule
ns-3~\cite{ns3} & \xmark & \cmark & \cmark & \cmark & \cmark \\
OMNeT++~\cite{omnetpp} & \xmark & \cmark & \cmark & \cmark & \cmark \\
SimGrid~\cite{simgrid,flowwifi2022} & \cmark & \cmark & \cmark & \xmark & \xmark \\
WRENCH~\cite{wrench} & \cmark & \multicolumn{4}{c}{via SimGrid models} \\
CloudSim~\cite{cloudsim} & \multicolumn{5}{c}{cloud resource/application models} \\
iFogSim~\cite{ifogsim} & \multicolumn{5}{c}{application graphs; abstract transport} \\
EdgeCloudSim~\cite{edgecloudsim} & \multicolumn{5}{c}{edge tasks; analytical WLAN delay} \\
\textbf{\ncsim{}} & \cmark & \cmark & \cmark & \cmark & \xmark \\
\bottomrule

\end{tabular}
\end{table}

\section{Architecture}
\label{sec:arch}

\begin{figure}[t]
\centering
\includegraphics[width=\columnwidth]{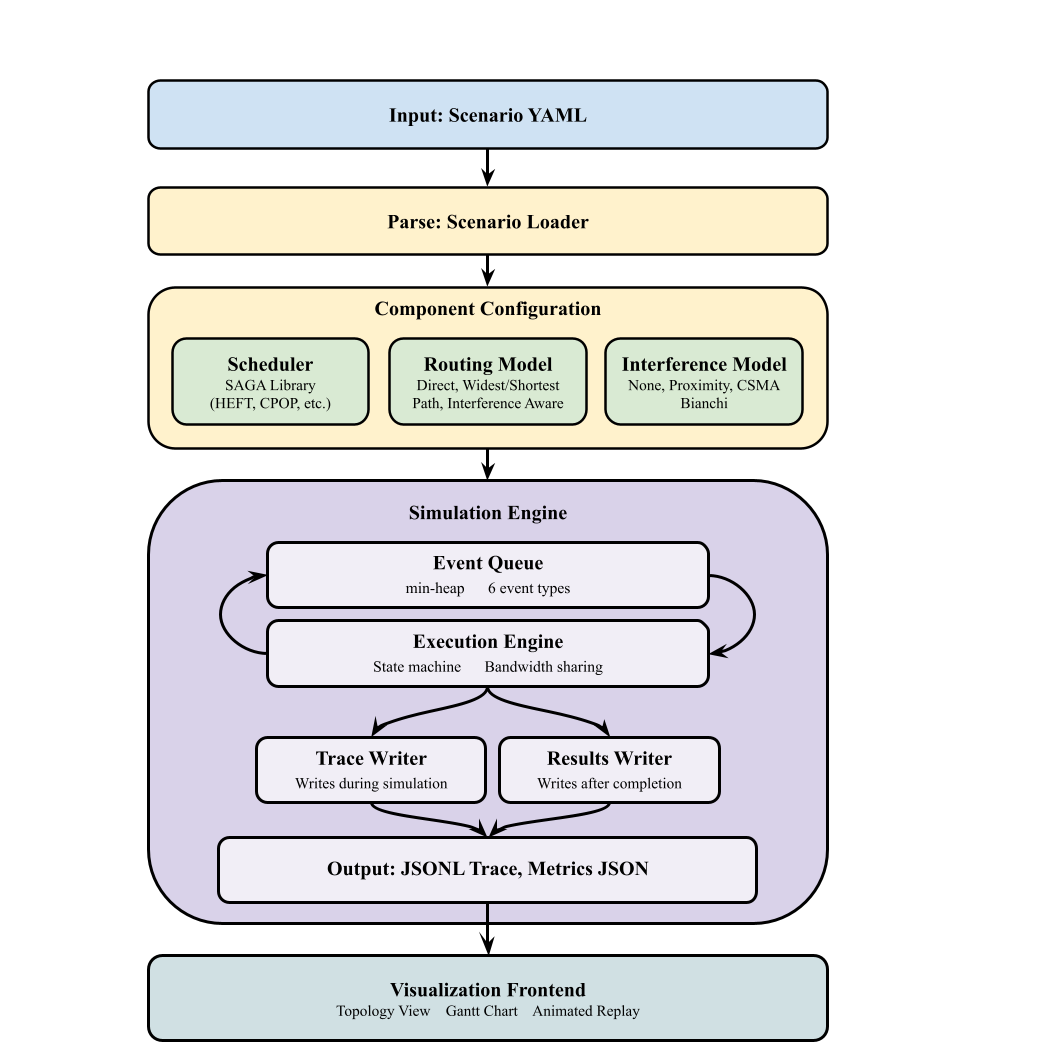}
\caption{High-level architecture of \ncsim{}.}
\label{fig:architecture}
\end{figure}

\ncsim{} is organized around a layered discrete-event simulation (DES) engine with pluggable
components for scheduling, routing, and interference modeling.
\Cref{fig:architecture} illustrates the data flow from scenario
specification through simulation to output traces.

\subsection{Design Goals}

Four principles guide \ncsim{}.
\textbf{Determinism:} stable event ordering, seeded inputs, and microsecond time quantization support repeatable runs in a recorded software environment.
\textbf{Modularity:} six abstract base classes (\Cref{tab:abstractions}) decouple schedulers, routing, and interference models so swapping one does not touch the others.
\textbf{Simplicity:} scenarios are declarative YAML; the simulator is a Python package with declarative inputs and pluggable models.
\textbf{Physical grounding:} link bandwidths are derived from RF parameters (transmit power, frequency, path loss exponent), so link rates are physically consistent with node positions rather than specified as arbitrary constants.

\subsection{Scope and Assumptions}

\ncsim{} targets macroscopic scheduling decisions using flow-level
wireless service and single-slot FIFO compute execution. It models CSMA/CA
contention, hidden terminals, and modulation and coding scheme (MCS)
selection in static networks.
\Cref{tab:scope} summarizes the modeling scope.

\begin{table}[t]
\centering
\caption{Modeling scope of \ncsim{}. The representation column identifies
the abstraction; validation results are reported in \Cref{sec:validation}.}
\label{tab:scope}
\begin{tabular}{@{}lll@{}}
\toprule
\textbf{Phenomenon} & \textbf{Representation} & \textbf{Scope} \\
\midrule
CSMA/CA contention & Bianchi & Local saturation \\
Hidden terminals & SINR & Persistent demand \\
Rate selection & 11n/ac/ax lookup & Heuristic thresholds \\
Path loss & Log-distance & Static geometry \\
Pkt-level dynamics & Not modeled & Unresolved \\
TCP slow-start & Not modeled & No transport stack \\
Static shadowing & Optional log-normal & Fixed realization \\
Multi-slot parallelism & Not modeled & One FIFO slot \\
\bottomrule
\end{tabular}
\end{table}

\ncsim{}'s flow-level abstraction targets scheduler evaluation with bulk DAG
edges (10\,MB-class payloads in the grid experiments), where sustained
service rates are useful approximations. It represents dependency payload
service rather than a TCP connection: no congestion window, retransmission
queue, or transport startup is simulated. The Bianchi component describes
homogeneous saturated contention, and its use in overlapping neighborhoods
is an approximation. We use deterministic path loss throughout; optional
log-normal shadowing is drawn once per node pair and held fixed. Mobility,
time-varying fading, and OFDM waveform dynamics are outside this implementation.

\subsection{Key Abstractions}

Six abstract base classes (\Cref{tab:abstractions}) let researchers plug
in their own schedulers, routing models, or interference models without
touching the core event engine; every component used in
\Cref{sec:eval} is itself an instance of these interfaces.

\begin{table}[t]
\centering
\caption{Pluggable abstractions in \ncsim{}.}
\label{tab:abstractions}
\begin{tabular}{@{}lp{4.8cm}@{}}
\toprule
\textbf{ABC} & \textbf{Implementations} \\
\midrule
Scheduler & Manual, RoundRobin, 22 SAGA algorithms; optional PEFT and structural baselines \\
RoutingModel & Direct, WidestPath, MinimumHop, ShortestPath \\
InterferenceModel & None, CSMA Bianchi \\
LinkModel & Static \\
QueueModel & FIFO \\
DAGSource & Single, Multi \\
\bottomrule
\end{tabular}
\end{table}

\subsection{Visualization and Tooling}
\label{sec:viz}

\ncsim{} includes a React/D3 web-based visualization frontend with three
views: a \textbf{network topology view} showing nodes, links, and active
transfers; a \textbf{Gantt chart} of task execution and transfer timelines
per node; and an \textbf{animated replay} with playback controls for
observing bandwidth recalculation cascades.  Simulation output is written
in JSONL trace format (one JSON object per event) for integration with
external analysis tools.  The command-line interface supports batch
execution, facilitating the automated parameter sweeps used throughout
\Cref{sec:eval}.

\subsection{Artifact Availability}
\label{sec:artifact}

The \ncsim{} project repository is
\url{https://github.com/ANRGUSC/ncsim}. Study materials are organized under
\path{artifacts/arxiv-2605.01094/} on the repository's \texttt{paper} branch
(\Cref{tab:artifact}). The folder contains the inputs, saved observations, analysis scripts, and
manuscript sources. A content-hash manifest identifies these files and the
simulator source used with them. Appendix~\ref{app:reproduction} gives the
folder layout and commands for checking the records, generating figures,
running the workflow studies, and compiling the paper.

\begin{table}[t]
\centering
\caption{Study artifact organization. Folder paths are relative to the
study directory; simulator source is at the repository root.}
\label{tab:artifact}
\begin{tabular}{@{}p{4.3cm}p{3.4cm}@{}}
\toprule
\textbf{Component} & \textbf{Location} \\
\midrule
\ncsim{} source identity & \path{MANIFEST.json} \\
Frozen workload and topology inputs & \path{inputs/} \\
Saved workflow and packet observations & \path{results/} \\
ns-3 sources and build instructions & \path{ns3/} \\
Validation, plotting, and analysis & \path{scripts/} \\
LaTeX source and figure dependencies & \path{manuscript/} \\
\bottomrule
\end{tabular}
\end{table}

\section{System Model}
\label{sec:model}

The models underlying \ncsim{} describe the
network and compute infrastructure, the DAG workflow representation,
the scheduling and routing frameworks, and the analytical
802.11 WiFi model that captures contention and hidden terminal effects.

\subsection{Network Model}

A network consists of \emph{nodes} and directed \emph{links}.  Each node $v$
has a compute capacity $C_v$ (compute units per second) and an optional 2D
position $(x_v, y_v)$ used for RF calculations.  Each link $\ell$ from node
$u$ to node $v$ has a bandwidth $B_\ell$ (MB/s) and latency $L_\ell$
(seconds).  Topologies are specified in YAML and can be arbitrary directed
graphs.

\subsection{Compute Model}

A task $t$ has a compute cost $w_t$ (in compute units).  The execution time
on node $v$ is $w_t / C_v$.  Each node processes one task at a time using
FIFO queuing.  Tasks follow a five-state lifecycle: \textsc{Pending}
$\rightarrow$ \textsc{Ready} $\rightarrow$ \textsc{Queued} $\rightarrow$
\textsc{Running} $\rightarrow$ \textsc{Completed}.

\subsection{DAG Workflow Model}

Workflows are modeled as directed acyclic graphs (DAGs) where nodes represent
tasks and edges represent data dependencies.  Each edge $e = (t_i, t_j)$
carries a data payload of size $d_e$ (in MB) that must be transferred from
the node executing $t_i$ to the node executing $t_j$ before $t_j$ can begin.
If $t_i$ and $t_j$ are placed on the same node, the transfer is
instantaneous (zero cost).  DAGs are injected at specified simulation times.

\subsection{Scheduling Framework}

The scheduler receives a \texttt{NetworkSnapshot}, an immutable view of
node capacities, link bandwidths, queue depths, and active transfers, and
returns a \texttt{PlacementPlan} mapping each task to a node.  \ncsim{}
includes built-in policies and an adapter for SAGA's static schedulers:

\begin{itemize}[itemsep=1pt,leftmargin=*]
\item \textbf{Manual:} Uses \texttt{pinned\_to} annotations from the YAML.
\item \textbf{RoundRobin:} Cycles through nodes, respecting pin constraints.
\item \textbf{SAGA:} The adapter exposes 22 static scheduling algorithms,
  including HEFT, CPOP, Min-Min, Max-Min, and Sufferage~\cite{heft,SAGARepo}.
  PEFT is additionally available when provided by the installed SAGA version.
  The virtual network uses route bottleneck goodput for reachable node pairs.
  The adapter retains task placements; execution uses ready-time FIFO,
  not the processor timelines returned by the scheduling algorithms.
  A small virtual bandwidth discourages unreachable pairs, while the executor
  reports any required dependency lacking a usable route as unroutable.
\end{itemize}
\noindent The experiments in \Cref{sec:eval} compare HEFT, CPOP, and
RoundRobin. The broader scheduler interface supports other comparisons
without changing the executor.

\subsection{Routing and Bandwidth Sharing}

Four routing models determine the path for data transfers:

\begin{itemize}[itemsep=1pt,leftmargin=*]
\item \textbf{Direct:} Only allows transfers over explicit single-hop links.
\item \textbf{WidestPath:} Modified Dijkstra maximizing bottleneck bandwidth.
\item \textbf{MinimumHop:} Breadth-first search minimizing hop count.
\item \textbf{ShortestPath:} Dijkstra minimizing configured latency.
  We use MinimumHop in the experiments because their link latencies are zero.
\end{itemize}

Multi-hop transfers use pipelined fluid semantics: all route links are
occupied during the byte-service phase, the service rate is their minimum
(bottleneck), and configured path latency is paid once before this phase.
Routes are selected from clean link goodputs and held fixed during a transfer.
For equal bottleneck rates, WidestPath retains the first predecessor found,
processing equal-priority nodes by node ID and outgoing links by link ID;
it has no secondary hop-count objective.
This is not hop-by-hop store-and-forward packet delivery. Same-node
dependencies incur zero communication time. When multiple transfers share
a link, each receives a fair share:

\begin{equation}
B_\text{per-flow}(\ell) = \frac{B_\ell \cdot f(\ell)}{N_\ell}
\label{eq:fair_share}
\end{equation}
where $f(\ell)$ is the interference factor and $N_\ell$ is the number of
concurrent flows on link $\ell$.  When a transfer starts or completes, all
affected transfers are recalculated and their completion events rescheduled.

\subsection{WiFi / 802.11 Model}
\label{sec:wifi}

The WiFi model replaces static link bandwidths with rates derived from RF
parameters, and models contention and hidden terminal interference using
path-loss and analytical service models. The physical-layer and MAC models
draw on standard 802.11 analysis: path loss and MCS selection are standard;
carrier-sensing range and conflict graph construction are adapted from the
literature; we integrate Bianchi's established contention analysis with a
flow-level effective-rate hidden-interference approximation.
The following subsections formalize these mechanisms.

\subsubsection{Physical Layer}

The received power at distance $d$ uses log-distance path loss:
\begin{align}
\text{PL}(d) &= \text{PL}(d_0) + 10\,n\,\log_{10}\!\left(\frac{d}{d_0}\right) \label{eq:pathloss} \\
P_\text{rx}(d) &= P_\text{tx} - \text{PL}(d) \label{eq:rxpower}
\end{align}
where $\text{PL}(d_0)$ is the Friis free-space loss at reference distance
$d_0 = 1$\,m, $n$ is the path loss exponent, and $P_\text{tx}$ is the
transmit power.  The signal-to-noise ratio is $\text{SNR} = P_\text{rx} - N_0$
(in dB), where $N_0$ is the noise floor.  The path loss exponent $n$ ranges from 2 (free-space) to $\geq 3.5$
(dense indoor).

The PHY data rate is selected from the appropriate 802.11 MCS table based
on the computed SNR:
\begin{equation}
R_\ell = \max\{r_k : \text{SNR}_\ell \geq \text{SNR}_{\min,k}\}
\label{eq:mcs}
\end{equation}
If no MCS threshold is met, $R_\ell=0$.
The SNR thresholds are configurable model parameters, not requirements
mandated by IEEE~802.11. This selects the highest-rate MCS whose threshold is met,
producing a discrete rate staircase, a key feature that continuous-rate
models fail to capture.

\subsubsection{Carrier Sensing and Conflict Graph}

The carrier sensing range $d_\text{CS}$ is the maximum distance at which a
transmission triggers the clear channel assessment (CCA) mechanism:
\begin{equation}
d_\text{CS} = d_0 \cdot 10^{\,\frac{P_\text{tx} - \theta_\text{CCA} - \text{PL}(d_0)}{10\,n}}
\label{eq:csrange}
\end{equation}

The conflict graph $G_C = (L, E_C)$ is defined over links.  Without RTS/CTS,
links sharing an endpoint always conflict; otherwise two links conflict
if either transmitter can sense any node of the other link.  With RTS/CTS, any node of one link sensing any node of the other
creates a conflict.

\subsubsection{CSMA Bianchi Model (\texttt{csma\_bianchi})}
\label{sec:bianchi}

The dynamic model separates two interference mechanisms.  Let $\mathcal{A}$
be the set of currently active links, $\mathcal{C}_\ell = \mathcal{A} \cap
\mathcal{N}(\ell)$ the active contending neighbors, and $\mathcal{H}_\ell =
(\mathcal{A} \setminus \mathcal{N}(\ell)) \setminus \{\ell\}$ the candidate
hidden interferers. Only candidates whose power at the receiver reaches
$-105$\,dBm enter the hidden-power sum. An active link is one used by an
unfinished backlogged transfer; every hop of a fluid route is active.
Multiple flows on the same link share service but do not increase the
link-based contender count.

\paragraph{Contention factor.}
The number of contending stations is $n = 1 + |\mathcal{C}_\ell|$.  Using
Bianchi's saturation throughput analysis~\cite{bianchi2000}, the MAC
efficiency $\eta(n,R)$ uses the coupled equations for transmission
probability $\tau$ and collision probability $p$:
\begin{align}
\tau &= \frac{2}{W_{\min}+1+W_{\min}p\sum_{k=0}^{m-1}(2p)^k} \label{eq:tau} \\
p &= 1 - (1 - \tau)^{n-1} \label{eq:p}
\end{align}
with minimum contention window $W_{\min} = 16$ and maximum backoff stage
$m = 6$.  We solve numerically via bisection
(\Cref{sec:external_val}) to obtain the equilibrium $(\tau^*, p^*)$.
Here $W_{\min}$ counts backoff choices ($\mathrm{CW}_{\min}+1$), and the
finite-sum form avoids an apparent singularity at $p=1/2$.
For positive PHY rate $R$, the payload efficiency depends on both contender count and rate:
\begin{equation}
\eta(n,R)=\frac{P_S L/R}{P_I\sigma+P_ST_S(R)+P_CT_C(R)}
\label{eq:eta}
\end{equation}
where $P_I=(1-\tau)^n$, $P_S=n\tau(1-\tau)^{n-1}$, $P_C=1-P_I-P_S$,
$L$ is payload bits, and $\sigma$ is the idle-slot duration. Rates in
Mb/s and times in microseconds make $L/R$ a payload duration. Each
contender receives $R\eta(n,R)/(8n)$ MB/s. The numerator counts payload
airtime; the denominator includes the full MAC exchange.
Appendix~\ref{app:timing} gives the rate-dependent frame timings.

\paragraph{Hidden terminal factor.}
Hidden terminals transmit simultaneously with link~$\ell$, causing
interference at the receiver.  The SINR is:
\begin{equation}
\text{SINR}_\ell = 10\log_{10}\!\left(\frac{P_\text{rx}^{(\text{lin})}}{N_0^{(\text{lin})} + \sum_{j \in \mathcal{H}_\ell} P_j^{(\text{lin})}}\right)
\label{eq:sinr}
\end{equation}
A frame's MCS is fixed during its reception. The flow model instead maps
persistent offered interference to an effective service rate, using
thresholds shifted by a margin:
\begin{equation}
\theta_{\mathrm{eff},k}=\theta_{\mathrm{select},k}-\Delta,
\qquad \Delta=5\ \mathrm{dB}.
\label{eq:capture}
\end{equation}
The margin is a phenomenological model parameter, chosen independently
of ns-3 receiver calibration. Capture analyses~\cite{daneshgaran2008,zorzi1994,hadzivelkov2002}
motivate the use of received-power ratios in this approximation.
Let $R_\ell^{\mathrm{eff}}$ be the highest rate meeting the shifted SINR
threshold, capped at the clean rate $R_\ell$. It is zero when no threshold
is met. With no active hidden interferer, the margin is bypassed and
$R_\ell^{\mathrm{eff}}=R_\ell$. Define
\begin{equation}
f_\mathrm{HT}(\ell)=R_\ell^{\mathrm{eff}}/R_\ell,\qquad R_\ell>0.
\label{eq:sinr_factor}
\end{equation}
Each active hidden link contributes its transmitter's received power to
the sum while the transfer has data remaining. The sum counts links:
two active outgoing links from the same transmitter contribute separately,
although a single radio sends its frames sequentially. This calculation
summarizes concurrent transfer demand without tracking when each radio
transmits or waits between packets. The optional fixed-MCS mode in
\Cref{sec:ns3_validation} instead estimates how often a hidden transmitter
overlaps a reception and reduces goodput when that overlap prevents
decoding. It applies to isolated pairs of hidden links and is used in
the two-link comparison in \Cref{fig:ns3_validation}.

\paragraph{Combined factor.}
The clean solo bandwidth used in \Cref{eq:fair_share} is
$B_\ell=R_\ell\eta(1,R_\ell)/8$ for $R_\ell>0$, and zero otherwise.
We set $f(\ell)=0$ when $R_\ell^{\mathrm{eff}}=0$.
For positive effective rate, the normalized interference factor is
\begin{equation}
\boxed{f(\ell)=
 \frac{R_\ell^{\mathrm{eff}}\eta(n,R_\ell^{\mathrm{eff}})}
 {nR_\ell\eta(1,R_\ell)}}
\label{eq:combined}
\end{equation}
so an isolated link has $f(\ell)=1$. The local Bianchi calculation treats
the contenders as homogeneous, using the target link's effective rate
to calculate MAC efficiency even when neighboring links have different rates.
No positive floor is imposed in the main experiments. Zero-rate transfers
pause and are reevaluated when the active set changes.

The Solo mode retains single-link MAC overhead and same-link flow sharing,
but disables concurrent-link wireless degradation. Full mode adds
\Cref{eq:combined}; both give the scheduler the same clean solo link rates.
The raw-PHY mode is available for diagnostics. The factorial comparison
uses Solo as its baseline so that both modes include single-link MAC overhead.

\section{Simulation Engine}
\label{sec:engine}

The discrete-event simulation engine coordinates
task execution, data transfer, and interference dynamics through an
event-driven architecture.

\subsection{Event Processing}

\ncsim{} uses an event-driven DES with a min-heap priority queue.  Six event
types are processed in priority order at each simulation time:

\begin{enumerate}[itemsep=1pt,leftmargin=*]
\item \textsc{DagInject}: Invokes scheduler, initializes task states
\item \textsc{TaskComplete}: Frees node, triggers output transfers
\item \textsc{TransferComplete}: Delivers data, checks readiness
\item \textsc{TaskReady}: Starts task or enqueues if node busy
\item \textsc{TaskStart}: Begins execution, schedules completion
\item \textsc{TransferStart}: Acquires link bandwidth, schedules completion
\end{enumerate}

The priority ordering at the same simulation time is chosen for causal
correctness: completions are processed before starts, and task
completions before transfer completions, so that a freed node is
immediately available to any task that becomes ready at the same
instant.  Stale events are cancelled via lazy deletion and discarded
when they reach the head of the queue; all times are rounded to
microsecond precision. Stable insertion sequence breaks
remaining ties; this specifies repeatability within the recorded environment.

\subsection{Bandwidth Recalculation Cascade}

Wireless interference means that the effective bandwidth of a link depends
on which other links are simultaneously active. When a transfer starts or
finishes, the engine collects same-link sharers and transfers affected through
carrier-sense or hidden-interference dependencies. It deduplicates these
flows and updates their progress before replacing their rates:
\begin{equation}
D_q(t)=\max\{0,D_q(t_0)-b_q(t_0)(t-t_0)\}.
\label{eq:progress}
\end{equation}
Here $D_q$ is remaining MB and $b_q(t_0)$ the previously assigned bottleneck
rate. Crediting elapsed service at the old rate preserves already-served
bytes. The engine then evaluates every hop of each affected route, takes
the new bottleneck, cancels obsolete completions, and schedules the remaining
work. Transfer completions are rounded upward so unserved bytes are not
discarded.

All hops remain active while their transfer is backlogged, including at zero
service. Collecting affected flows once suffices because rate changes
do not change which links remain active.
A subsequent start or finish triggers another expansion. Dense hidden
neighborhoods can make this nearly global; a small conflict-graph degree
does not bound the work. Heap operations cost $O(\log Q)$ for $Q$ pending
events, while refresh cost also depends on affected routes and neighborhoods.

FIFO compute execution reserves the next queued task's node before its
start event is processed, preventing two equal-time starts from claiming
one processor. A paused transfer can resume after competing traffic ends.
If the event queue empties with unfinished wireless backlogs, the run is
reported as \texttt{blocked\_wireless}, a deadlock of this deterministic
offered-interference model. It is not a proof of physical infeasibility.
Required dependencies with no usable route are \texttt{unroutable}; explicit
simulation limits and implementation errors have separate statuses.
Only completed runs receive a finite makespan. Appendix~\ref{app:engine}
gives a worked two-hop example and the event-update checks.

\section{Validation}
\label{sec:validation}

We evaluate \ncsim{}'s flow-level model in three validation stages.
\textbf{Stage~1 (Internal).} Controlled experiments check link rates,
contention, hidden-interference composition, and event updates against
their analytical definitions.
\textbf{Stage~2 (Literature).} We reproduce Bianchi's Table~III and
Figure~6 to check the MAC analytical component~\cite{bianchi2000}.
\textbf{Stage~3 (Packet comparison).} We compare with fixed-rate
ns-3 contention and separation experiments and distinguish the approximation
used for each curve. These stages test implementation consistency,
reproduction of a published analytical model, and predictive agreement
in specified packet configurations, respectively.

\subsection{Internal Verification}
\label{sec:internal_val}

The internal experiments isolate link-rate selection, parallel
interference, shared receivers, staggered starts, same-link sharing,
N-way contention, hidden-interference cascades, and combined effects.
The numerical examples below exercise the default
model; the event-engine regression suite also checks recovery
from temporary zero rates, FIFO execution, and local versus global refresh.
These checks establish consistency between the model definitions and execution.

\subsubsection{Experiment 1: Link Length vs.\ Data Rate}

\begin{figure}[t]
\centering
\includegraphics[width=\columnwidth]{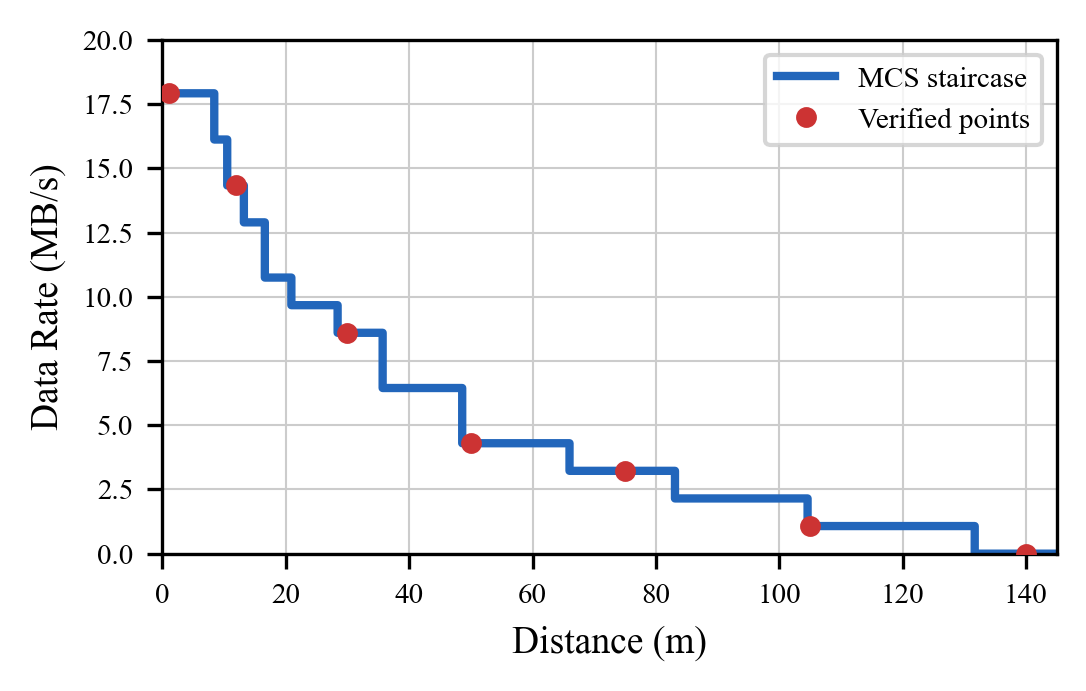}
\caption{Experiment 1: PHY data rate vs.\ link distance showing discrete MCS
  transitions (802.11ax, 5\,GHz, $n{=}3$).  Red dots mark experimentally
  verified data points from \Cref{tab:exp1}.}
\label{fig:exp1}
\end{figure}

A single link between two nodes at varying distances from 1\,m to 140\,m.
\Cref{tab:exp1} shows selected results.  The simulated rate matches the
analytically predicted MCS-selected rate at every distance.  At 140\,m the
SNR (4.19\,dB) falls below MCS~0 (5\,dB threshold), correctly yielding zero
rate.  \Cref{fig:exp1} plots the characteristic staircase pattern: rate
is constant within each MCS band and drops discretely at each SNR threshold.

\begin{table}[t]
\centering
\caption{Experiment 1: PHY-rate verification versus link length
(802.11ax, 5\,GHz). Rates exclude MAC overhead.}
\label{tab:exp1}
\begin{tabular}{@{}rrrrr@{}}
\toprule
\textbf{Dist(m)} & \textbf{SNR(dB)} & \textbf{MCS} & \textbf{Pred(MB/s)} & \textbf{Sim(MB/s)} \\
\midrule
1   & 68.58 & 11  & 17.925 & 17.925 \\
12  & 36.20 & 9   & 14.338 & 14.338 \\
30  & 24.27 & 5   & 8.600  & 8.600  \\
50  & 17.61 & 3   & 4.300  & 4.300  \\
75  & 12.33 & 2   & 3.225  & 3.225  \\
105 & 7.94  & 0   & 1.075  & 1.075  \\
140 & 4.19  & n/a & 0.000  & 0.000  \\
\bottomrule
\end{tabular}
\end{table}

\subsubsection{Experiment 2: Parallel Link Separation}

\begin{figure}[t]
\centering
\includegraphics[width=\columnwidth]{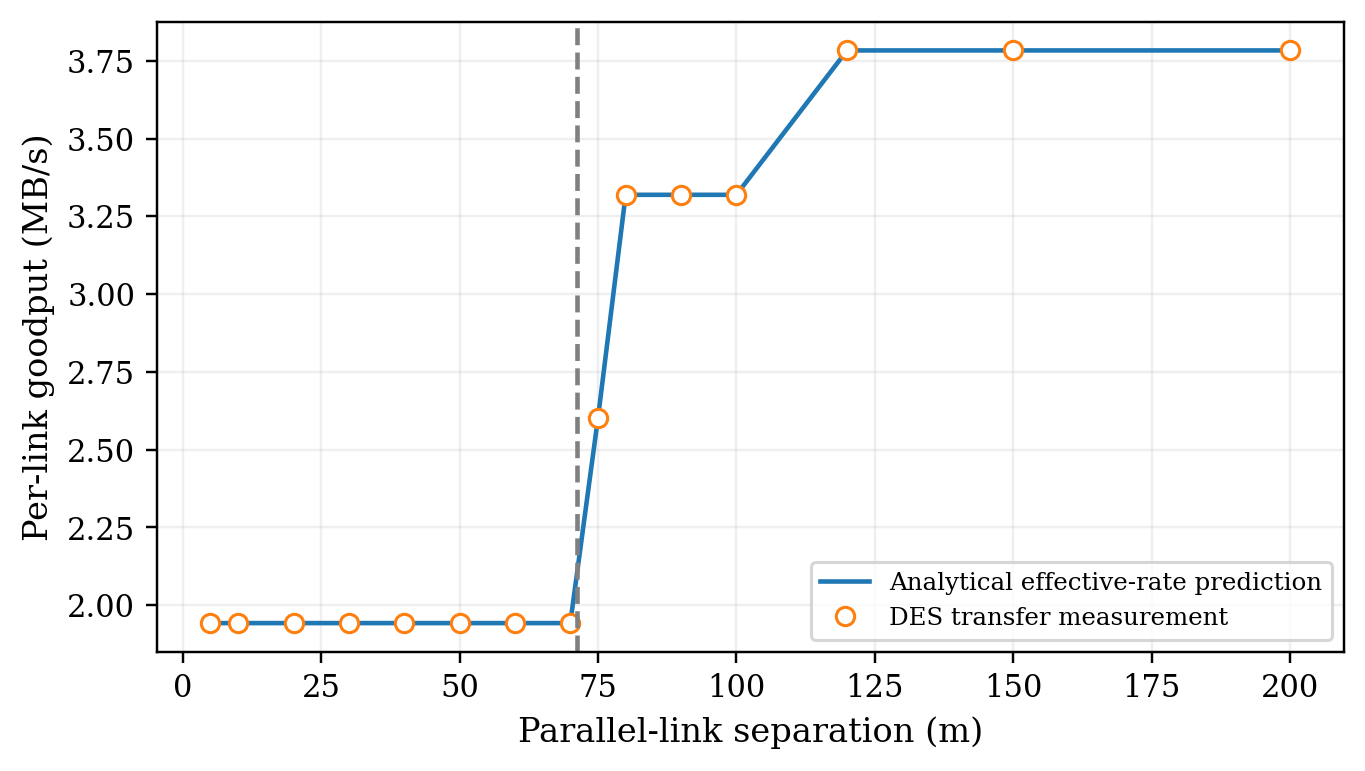}
\caption{Experiment 2: effective per-link rate vs.\ separation for two
  parallel 30\,m links. The sensing boundary is 71.2\,m. Points are DES
  transfer measurements and the line is the analytical prediction.}
\label{fig:exp2}
\end{figure}

Two parallel 30\,m links separated vertically by distances from 5\,m to
200\,m.  \Cref{fig:exp2} illustrates the three interference regimes.
At separation $\leq70$\,m, the transmitters are within the carrier-sensing
range (71.2\,m), so the links contend: each receives
$8.6\,\eta(2,68.8)/2=1.942$\,MB/s. The clean PHY rate is 68.8\,Mb/s
and the solo payload goodput is 3.783\,MB/s.

Beyond the sensing boundary, the effective-rate model assigns
2.601\,MB/s at 75\,m, 3.319\,MB/s at 80--100\,m, and 3.783\,MB/s from
120\,m in the tested sweep. All 15 transfer measurements agree with the
piecewise analytical calculation to the reported precision.
\Cref{fig:ns3_validation} compares this effective-rate prediction with ns-3
and with the optional fixed-capture mode, which models the hidden-terminal
dip through packet overlap.

\subsubsection{Experiment 4: Three Parallel Links}

This experiment extends the two-link topology of Experiment~2 to three
parallel 30\,m links placed at $y = 0$, $y = s$, and $y = 2s$ (links A,
B, C), each carrying a 10\,MB transfer.  \Cref{fig:exp4_setup} illustrates
the topology with separation~$s$ as the independent variable.

\begin{figure}[t]
\centering
\begin{tikzpicture}[
  nd/.style={circle, fill=black, inner sep=0pt, minimum size=5pt},
  lnk/.style={->, thick, blue!70!black},
  sep/.style={<->, densely dashed, gray!70!black},
  annot/.style={font=\scriptsize, text=gray!70!black},
]
  \node[nd, label={[font=\scriptsize]left:$n_0$}] (a0) at (0,0) {};
  \node[nd, label={[font=\scriptsize]right:$n_1$}] (a1) at (3.2,0) {};
  \draw[lnk] (a0) -- node[above, font=\scriptsize] {Link A} (a1);

  \node[nd, label={[font=\scriptsize]left:$n_2$}] (b0) at (0,1.3) {};
  \node[nd, label={[font=\scriptsize]right:$n_3$}] (b1) at (3.2,1.3) {};
  \draw[lnk] (b0) -- node[above, font=\scriptsize] {Link B} (b1);

  \node[nd, label={[font=\scriptsize]left:$n_4$}] (c0) at (0,2.6) {};
  \node[nd, label={[font=\scriptsize]right:$n_5$}] (c1) at (3.2,2.6) {};
  \draw[lnk] (c0) -- node[above, font=\scriptsize] {Link C} (c1);

  \draw[sep] (-0.55,0) -- node[left, annot] {$s$} (-0.55,1.3);
  \draw[sep] (-0.55,1.3) -- node[left, annot] {$s$} (-0.55,2.6);

  \draw[sep] (3.75,0) -- node[right, annot] {$2s$} (3.75,2.6);

  \draw[<->, gray!70!black] (0,-0.4) -- node[below, annot] {30\,m} (3.2,-0.4);

  \node[annot, align=left, anchor=west] at (4.8,1.3) {%
    $s \leq 35.6$\,m: all conflict\\
    $35.6 < s \leq 71.2$\,m: mixed\\
    $s > 71.2$\,m: all hidden};
\end{tikzpicture}
\caption{Experiment~4 topology: three parallel 30\,m links at variable
  separation~$s$.  As $s$ increases, the interference regime transitions
  from all-conflict to mixed (adjacent pairs contend, outer pair A--C
  are hidden terminals) to all-hidden.}
\label{fig:exp4_setup}
\end{figure}

As $s$ increases, three interference regimes emerge: (i)~\emph{all-conflict}
($s \leq 35.6\,$m): all three pairs are within $d_{\text{CS}}$, each link
gets $8.6\,\eta(3,68.8)/3$\,MB/s; (ii)~\emph{mixed}
($35.6 < s \leq 71.2\,$m): adjacent pairs (A--B, B--C) contend via
Bianchi, while the outer pair (A--C at distance~$2s$) are hidden terminals,
creating asymmetric rates; and (iii)~\emph{all-hidden} ($s > 71.2\,$m):
no conflict-graph edges, each link sees the others purely as hidden
terminals.  Because A/C and B generally have different rates, one group
finishes first, and the remaining links' interference conditions change.
The analytical prediction accounts for this multi-phase behavior.

\begin{table}[t]
\centering
\caption{Experiment~4: Three parallel links at varying separation.
  In the mixed regime the middle link~B contends with both neighbors
  while outer links A, C contend only with B and see each other as
  hidden terminals.}
\label{tab:exp4}
\begin{tabular}{@{}rlrrrr@{}}
\toprule
$s$(m) & Regime & Pred $R_A$ & Sim $R_A$ & Pred $R_B$ & Sim $R_B$ \\
\midrule
10 & all-conf & 1.275 & 1.275 & 1.275 & 1.275 \\
35 & all-conf & 1.275 & 1.275 & 1.275 & 1.275 \\
40 & mixed & 1.686 & 1.686 & 1.521 & 1.521 \\
50 & mixed & 1.686 & 1.686 & 1.521 & 1.521 \\
70 & mixed & 1.942 & 1.942 & 1.651 & 1.651 \\
75 & all-hid & 2.601 & 2.601 & 2.601 & 2.601 \\
100 & all-hid & 3.319 & 3.319 & 2.790 & 2.790 \\
150 & all-hid & 3.783 & 3.783 & 3.783 & 3.783 \\
\bottomrule
\multicolumn{6}{@{}l}{\scriptsize A$=$C by symmetry in all cases.  Rates in MB/s.}
\end{tabular}
\end{table}

\Cref{tab:exp4} shows transfer-average rates, including rate changes after
one group finishes. All 11 separation cases agree within $4\,\mu$s in
completion time. In the mixed regime, the outer links finish before B:
at $s=70$\,m their average rate is 1.942\,MB/s versus B's 1.651\,MB/s.
At 100\,m the corresponding rates are 3.319 and 2.790\,MB/s. This checks
composition of contention and hidden-power effects with asymmetric
completion events.

\subsubsection{Additional Validation Experiments}

Five additional experiments validate: (3)~two-transmitter shared-receiver
topologies, (5)~staggered transfer starts with dynamic
recalculation, (6)~per-link bandwidth sharing with multiple flows,
(8)~five-link hidden terminal cascades with different data sizes, and
(9)~combined bandwidth sharing with hidden terminal interference.  The regression suite covers these interaction types; the numerical
parallel-link checks are also executed by the evidence-generation script.

\subsection{External Validation Against Bianchi (2000)}
\label{sec:external_val}

To check our implementation against an external reference, we
reproduce the numerical results from Bianchi's paper
\cite{bianchi2000} using his exact parameters: FHSS 1\,Mbps basic
access (no RTS/CTS), with slot duration 50\,$\mu$s, SIFS 28\,$\mu$s,
DIFS 128\,$\mu$s, propagation delay 1\,$\mu$s, payload 8184 bits,
MAC header 272 bits, PHY header 128 bits, and ACK 112 bits plus the PHY header.

\subsubsection{Table III Reproduction}

\Cref{tab:bianchi_table3} compares our computed normalized saturation
throughput $S$ against the values published in Bianchi's Table~III for
$W{=}32$, $m{=}3$.  At $n{=}2$, our solver computes $S = 0.847311$
versus the published $0.8473$ (relative error $0.001\%$).  At $n{=}3$,
we compute $S = 0.836828$ versus $0.8368$ (relative error $0.003\%$).
These sub-basis-point differences are consistent with rounding in the
published table.

\begin{table}[t]
\centering
\caption{Reproduction of Bianchi~\cite{bianchi2000} Table~III:
normalized saturation throughput $S$ for $W{=}32$, $m{=}3$.}
\label{tab:bianchi_table3}
\begin{tabular}{@{}rrrr@{}}
\toprule
$n$ & \textbf{Computed} & \textbf{Published} & \textbf{Rel.\ Error} \\
\midrule
2  & 0.847311 & 0.8473 & 0.001\% \\
3  & 0.836828 & 0.8368 & 0.003\% \\
\bottomrule
\end{tabular}
\end{table}

We solve the coupled equations (\Cref{eq:tau,eq:p}) by bisection rather
than direct fixed-point iteration, which can oscillate when the mapping slope
exceeds unity. The solver brackets $p$ on $[0,1]$, checks an equation
residual of at most $10^{-13}$, and raises an error if 100 steps do not
satisfy that check.

\subsubsection{Figure 6 Reproduction}

\begin{figure}[t]
\centering
\includegraphics[width=\columnwidth]{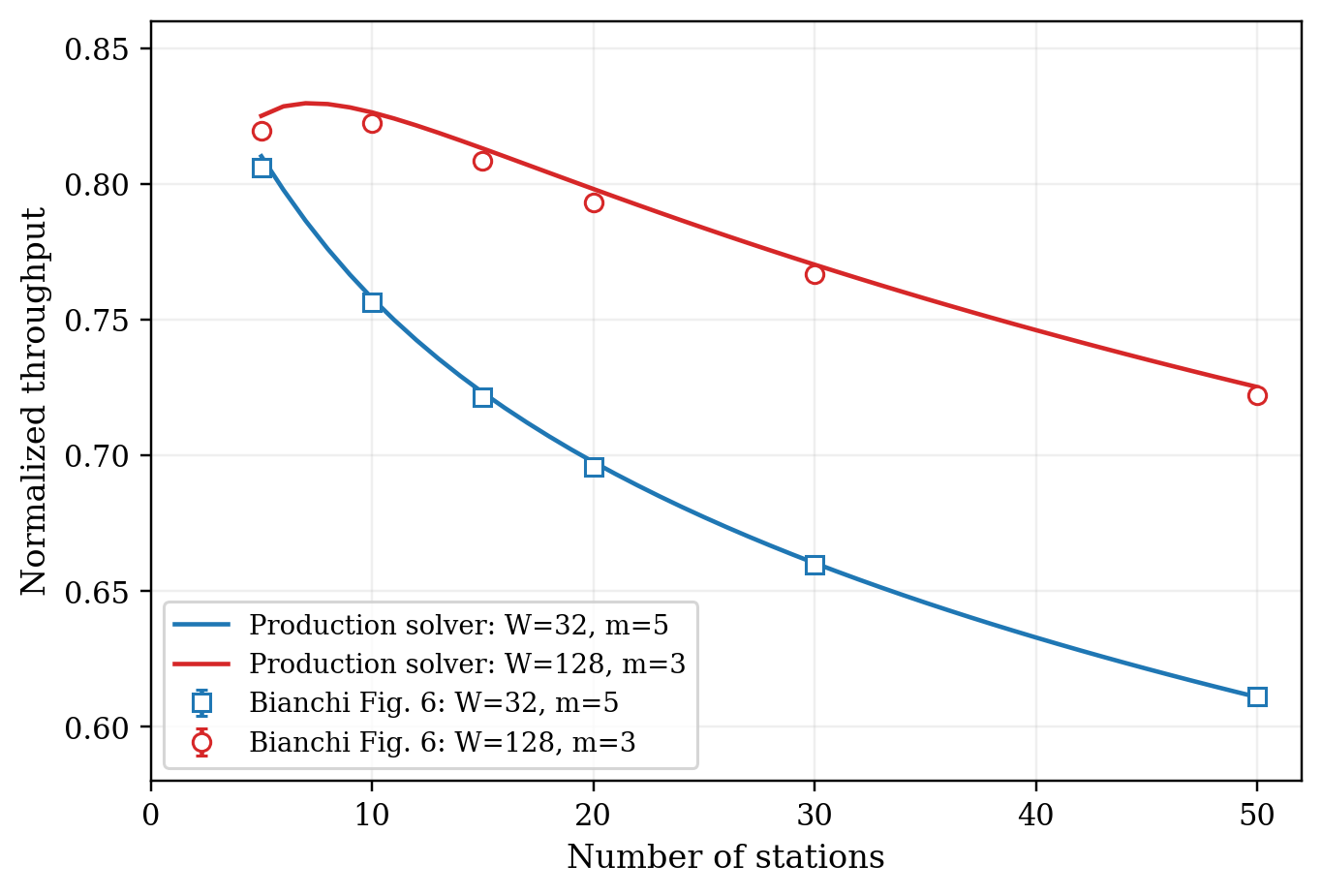}
\caption{Reproduction of Bianchi~\cite{bianchi2000} Figure~6:
  normalized saturation throughput vs.\ number of stations for two
  $(W, m)$ configurations.  The $W{=}128$ curve's non-monotonicity
  between $n{=}5$ and $n{=}10$ is consistent with Bianchi's published
  results.  Twelve digitized markers are shown with two-pixel reading uncertainty;
  their maximum relative difference from the solver is 0.66\%.}
\label{fig:bianchi_fig6}
\end{figure}

\Cref{fig:bianchi_fig6} reproduces Bianchi's Figure~6 throughput curves
for $W{=}32$, $m{=}5$ and $W{=}128$, $m{=}3$ across $n{=}5$ to $n{=}50$
stations.  The smaller-window curve decreases over this range; the $W{=}128$ curve reproduces the slight non-monotonicity between
$n{=}5$ and $n{=}10$ consistent with Bianchi's published results.
Twelve marker centers read from Bianchi's published figure differ from our
computed curves by at most 0.66\%. The archived image, axis calibration,
and marker coordinates are retained; two-pixel reading uncertainty is
approximately $0.0017$ in normalized throughput. This graphical check has
lower precision than the numerical Table~III comparison.

\subsubsection{N-Way Contention Scaling}

To confirm that the Bianchi model scales correctly within the full
simulation loop, we place $n \in \{2, \ldots, 8\}$ parallel 30\,m links
at 5\,m separation (all within carrier-sensing range), each carrying a
10\,MB transfer.

\begin{table}[t]
\centering
\caption{N-way Bianchi contention scaling (30\,m links at 5\,m
separation).  $\eta(n,68.8)$ is the payload efficiency at 68.8\,Mb/s.}
\label{tab:exp7}
\begin{tabular}{@{}rrrrrr@{}}
\toprule
$n$ & $\eta(n)$ & $\eta(n)/n$ & \textbf{Pred(MB/s)} & \textbf{Sim(MB/s)} \\
\midrule
2 & 0.452 & 0.226 & 1.942 & 1.942 \\
3 & 0.445 & 0.148 & 1.275 & 1.275 \\
4 & 0.435 & 0.109 & 0.936 & 0.936 \\
5 & 0.427 & 0.085 & 0.734 & 0.734 \\
6 & 0.419 & 0.070 & 0.601 & 0.601 \\
7 & 0.413 & 0.059 & 0.507 & 0.507 \\
8 & 0.407 & 0.051 & 0.437 & 0.437 \\
\bottomrule
\end{tabular}
\end{table}

\Cref{tab:exp7} confirms the Bianchi solver produces correct throughput
predictions when embedded in the full simulation engine with real RF
parameters and dynamic recalculation.

\subsection{Cross-Validation Against ns-3}
\label{sec:ns3_validation}

Stages~1 and~2 establish internal consistency and reproduction of Bianchi's
analytical framework. Stage~3 compares predicted goodput with ns-3.41
packet-level simulation. Two fixed-MCS experiments test homogeneous
contention and the transition to hidden interference. We identify the
service approximation used for each predicted curve.

Both experiments use 802.11ax at 5\,GHz on a 20\,MHz channel, one spatial
stream, \texttt{ConstantRateWifiManager} with \texttt{HeMcs5}, and the
settings in \Cref{tab:ns3_params}. A-MPDU and A-MSDU are disabled.
Each source offers saturated UDP traffic with 1472-byte application payloads.
The measurement is total received application payload divided by 29.5\,s:
traffic starts at 0.5\,s and stops at 30\,s, with no additional warmup
discard. Each point first averages link goodput within each of 20 seeds,
then reports the mean and a Student-$t$ 95\% confidence interval across seeds.
This avoids treating correlated links within one run as independent samples.

\subsubsection{Experiment 1: Contention Scaling}

We place $n \in \{1, \ldots, 8\}$ parallel STA-AP link pairs at 30\,m
link length and 5\,m vertical separation, all well within the 71.2\,m
carrier-sensing range, creating a complete conflict graph.  Each STA sends
saturated UDP traffic to its AP.  \Cref{fig:ns3_validation}(a) overlays
\ncsim{}'s Bianchi prediction $8.6\,\eta(n,68.8)/n$\,MB/s against the
ns-3 per-station goodput.

\ncsim{}'s Bianchi model accounts for preamble, SIFS, ACK, and inter-frame
overhead in the payload efficiency. It tracks ns-3 closely across the
station counts: the isolated-link prediction is 3.783\,MB/s versus
3.770\,MB/s in ns-3, a 0.35\% difference. Across the eight settings, mean
absolute relative error is 5.1\% and the maximum is 8.2\% at seven links.
Here error is $|\widehat g-\overline g_{\mathrm{ns3}}|/
\overline g_{\mathrm{ns3}}$, giving each station-count setting equal weight.
The seed-level confidence half-widths are at most 0.0025\,MB/s. The
decreasing curves and small seed variability show consistent agreement
across the tested homogeneous contention settings.

\subsubsection{Experiment 2: Separation Sweep}

Two parallel 30\,m links are placed at 16 separations:
10, 20, 30, 40, 50, 60, 65, 70, 72, 75, 80, 90, 100, 120, 150, and
200\,m. Beyond the 71.2\,m sensing boundary, simultaneous transmissions
can corrupt frames at the receiver.

The optional fixed-MCS mode uses a \emph{capture/overlap
approximation}. In the hidden regime it multiplies the solo goodput by
a success probability: one if SINR meets the MCS~5 capture threshold,
otherwise the modeled idle fraction $1-a$, where $a$ is the interferer's
saturated data-airtime fraction. Equivalently, the success factor is
$(1-a)+a\,\mathbf{1}\{\mathrm{SINR}\geq\theta_{\mathrm{decode}}\}$.
With a
5\,dB margin this predicts 1.50\,MB/s just beyond the sensing boundary
and recovery to 3.78\,MB/s at 120\,m. The event engine executes this rule
when the fixed-capture option is selected. It holds the clean MCS fixed
and uses isolated saturated airtime as an overlap approximation; it does
not resolve packet retries or ACK interference. The option supports
isolated hidden-link pairs at 20\,MHz with RTS/CTS disabled and rejects
multiple hidden interferers or mixed hidden and sensed contention.

\begin{figure}[t]
\centering
\includegraphics[width=\columnwidth]{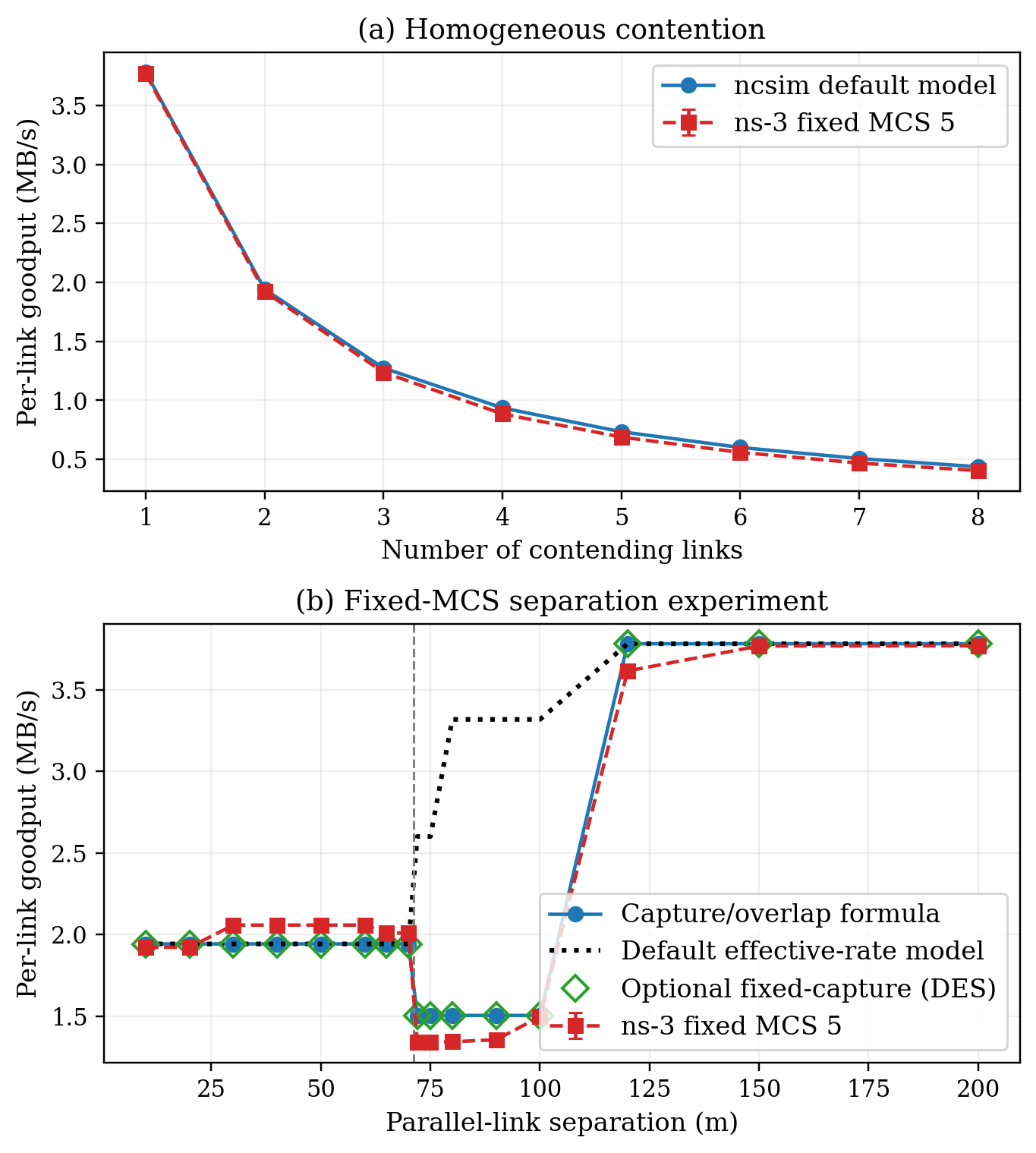}
\caption{ns-3 validation (20-seed means and 95\% CIs).
(a)~Homogeneous contention. (b)~The optional ncsim fixed-capture
mode (diamonds) implements the overlap approximation for two links.
Its mean and maximum absolute relative errors across 16 separations are
5.3\% and 12.2\%. The dotted curve uses the default effective-rate model,
which is also used in the workflow experiments.}

\label{fig:ns3_validation}
\end{figure}

\Cref{fig:ns3_validation}(b) shows this comparison. In the fixed-MCS
ns-3 data, per-link means near the hidden boundary are approximately
1.34--1.50\,MB/s, recovering to 3.62\,MB/s at 120\,m and 3.77\,MB/s
at 200\,m. The fixed-capture approximation reproduces this dip and recovery.
Across the sixteen settings, its mean and maximum absolute relative
errors are 5.3\% and 12.2\%, respectively, using the packet mean as
denominator. The figure also shows the default effective-rate prediction: 3.319\,MB/s
at 80\,m versus the packet mean 1.343\,MB/s, a 2.47-fold ratio.
At 120\,m the default-model prediction is 3.783\,MB/s versus 3.616\,MB/s,
a 4.6\% difference. The two rules therefore have different accuracy
near the sensing boundary. All workflow experiments use the default
effective-rate rule; the fixed-capture results concern this packet comparison.

\begin{table}[t]
\centering
\caption{Aligned parameters for ns-3 cross-validation.}
\label{tab:ns3_params}
\begin{tabular}{@{}lll@{}}
\toprule
\textbf{Parameter} & \textbf{ncsim} & \textbf{ns-3} \\
\midrule
TX power & 20\,dBm & \texttt{TxPowerStart/End = 20} \\
Frequency & 5\,GHz & Band 5\,GHz, ch.\,36 \\
Channel width & 20\,MHz & \texttt{ChannelSettings} \\
Path loss exp. & 3.0 & \texttt{Exponent = 3.0} \\
Ref.\ loss (1\,m) & 46.4\,dB & \texttt{ReferenceLoss = 46.4} \\
Noise floor & $-95$\,dBm & \texttt{RxNoiseFigure = 6} \\
CCA threshold & $-82$\,dBm & \texttt{CcaEdThreshold = -82} \\
CW$_{\min}$ & 16 & \texttt{MinCw = 15} \\
CW$_{\max}$ & 1024 & \texttt{MaxCw = 1023} \\
A-MPDU & disabled & \texttt{MaxAmpduSize = 0} \\
Guard interval & 800\,ns & \texttt{GuardInterval = 800} \\
\bottomrule
\end{tabular}
\end{table}

The ns-3 simulation scripts, Dockerfile for reproducible builds, and the
raw results are included in the evidence artifact, with source hashes
and measurement settings (Appendix~\ref{app:reproduction}).

\section{Experimental Evaluation}
\label{sec:eval}

The central question is whether concurrent wireless interference changes
scheduler choices. We examine
routing interactions (\Cref{sec:routing_eval}), absolute interference
impact (\Cref{sec:interference_eval}), the 108-run factorial
(\Cref{sec:scheduler_eval}), and sensitivity to payload size and concurrent
workflows (\Cref{sec:sensitivity}). All numerical results in this section
use the default effective-rate model and the execution semantics in
\Cref{sec:engine}.

The Solo baseline includes clean single-link MAC overhead. Both Solo
and Full use the same planning bandwidths and placements, so the paired
comparison isolates the effect of concurrent wireless service. Minimum-hop
routing minimizes hop count on the zero-latency links. The executor credits
progress at the assigned rate and enforces FIFO compute capacity.

All experiments use 802.11ax at 5\,GHz with a path loss exponent of
$n = 3$ (except where explicitly varied), heterogeneous compute capacities
(80--300 units/s), 10\,MB data payloads per DAG edge, and deterministic
seed~42. The grid capacities cycle through 200, 100, 150, 80, 300, 120,
250, 180, 160, 90, 220, 140, 280, 110, 190, and 170 units/s. Horizontal,
vertical, and checkerboard diagonal links form bidirectional meshes at
40\,m spacing.

Three DAG structures of increasing complexity are used across all
experiments:
\begin{itemize}[itemsep=1pt,leftmargin=*]
\item \textbf{Small (5 tasks):} Fork-join, $T_0 \to \{T_1, T_2, T_3\}
  \to T_4$.
\item \textbf{Medium (10 tasks):} Diamond, $T_0 \to \{T_1\text{--}T_4\}
  \to \{T_5\text{--}T_8\} \to T_9$ with selective cross-links between
  layers.
\item \textbf{Large (20 tasks):} Multi-level pipeline, $T_0 \to 4 \to 6
  \to 6 \to 3$ tasks per layer with sparse inter-layer connections.
\end{itemize}
\noindent All tasks have \texttt{compute\_cost}${}= 500$ and all edges have
\texttt{data\_size}${}= 10$\,MB (except where varied in
\Cref{sec:sensitivity}).

\subsection{Scheduler Information and Fairness}
\label{sec:scheduler_info}

Both the interference-free and interference-aware modes present an
identical \texttt{NetworkSnapshot} to the scheduler, containing the same
clean solo-goodput link bandwidths (\Cref{tab:scheduler_info}).  No scheduler
sees the conflict graph or contention factors; these are runtime phenomena
computed \emph{after} the schedule has been committed.  Any difference
in makespan is therefore attributable to the change in runtime wireless service under a fixed placement. The
SAGA algorithms supply placements and the common executor supplies FIFO
ordering; the comparison therefore concerns these placement adaptations.

\begin{table}[t]
\centering
\caption{Scheduler inputs.  No scheduler observes interference.}
\label{tab:scheduler_info}
\setlength{\tabcolsep}{3pt}
\begin{tabular}{@{}llll@{}}
\toprule
\textbf{Scheduler} & \textbf{Planner Inputs} &
  \textbf{Network View} & \textbf{Intf?} \\
\midrule
Manual     & Pin annotations      & N/A                & No \\
RoundRobin & Node list            & N/A                & No \\
HEFT       & DAG, $C_v$, $B_\ell$ & Full virtual net$^\dagger$ & No \\
CPOP       & DAG, $C_v$, $B_\ell$ & Full virtual net$^\dagger$ & No \\
\bottomrule
\multicolumn{4}{@{}l}{\scriptsize $^\dagger$Route bottleneck solo goodput for reachable pairs;
  0.001\,MB/s for unreachable.}
\end{tabular}
\end{table}

\subsection{Routing Strategy Comparison}
\label{sec:routing_eval}

Because multi-hop routing determines contention domains, we first
establish how routing strategies interact with interference before
evaluating schedulers.  We compare widest-path and minimum-hop routing
under HEFT scheduling with the interference-aware model across grid mesh
networks of three sizes (2$\times$2, 3$\times$3, 4$\times$4) and three
DAG complexities (5, 10, and 20 tasks).  All links use 40\,m grid
spacing with heterogeneous compute capacities (80--300 units/s).

\begin{figure}[t]
\centering
\includegraphics[width=\columnwidth]{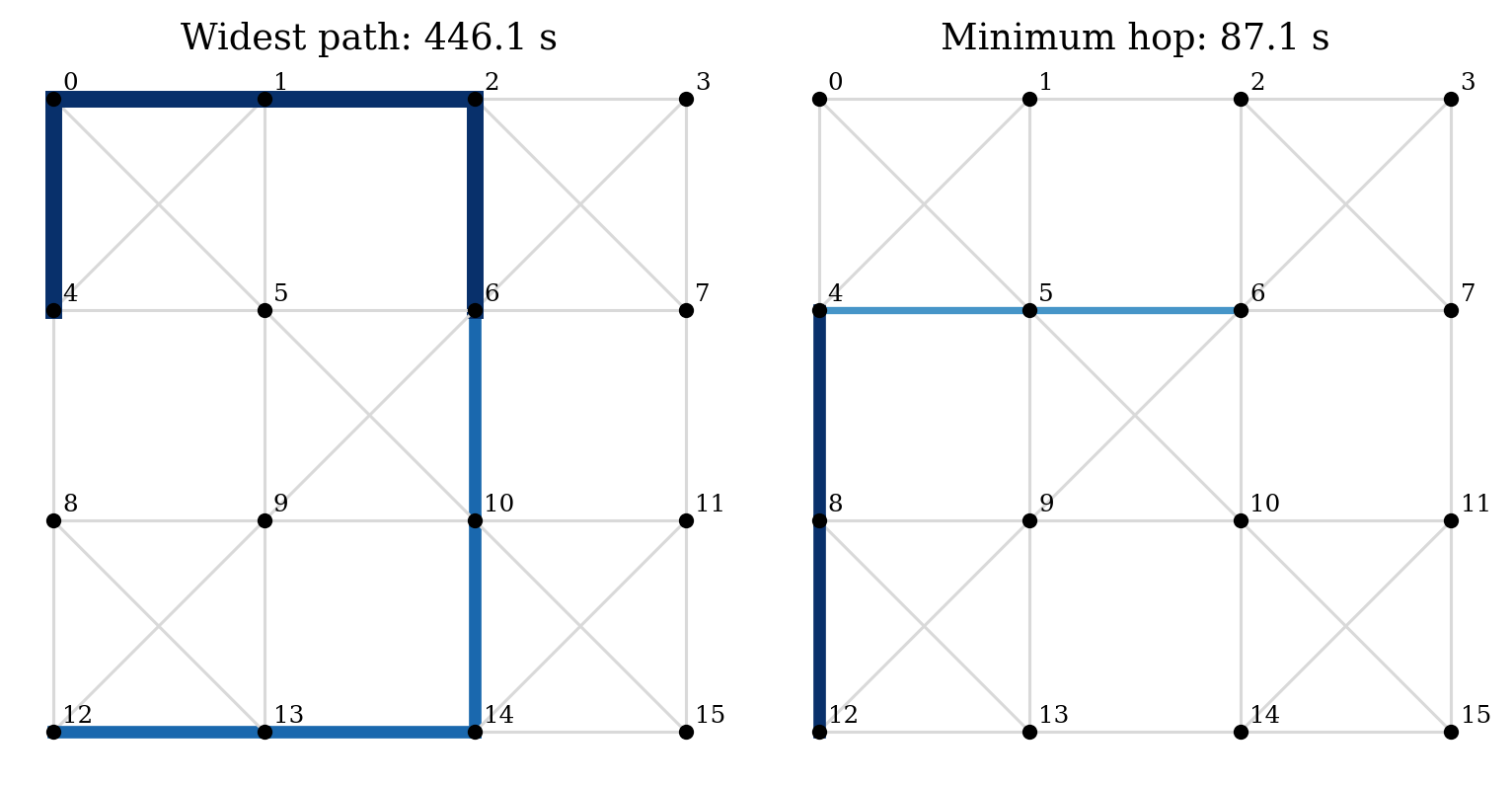}
\caption{Route load under widest-path (left) and minimum-hop (right) routing
on the 4$\times$4 grid with 20 tasks. Thickness and color show
routed dependency counts. Cross-node payload is 130\,MB in both cases;
route work is 880 versus 260\,MB-hops and Full makespan is 446.10
versus 87.07\,s.}

\label{fig:routing-topo}
\end{figure}

\Cref{fig:routing-topo} shows route load for the 4$\times$4, 20-task case.
Both routing configurations send 130\,MB across nodes. Widest-path routing uses
880\,MB-hops, averaging 6.77 hops per remote dependency, whereas minimum-hop
routing uses 260\,MB-hops and averages two hops. The former loads eight
undirected links, up to 13 routed dependencies per link; the latter uses
four links, up to nine dependencies per link. These counts describe the
dependencies routed through each link.

Under Full service the makespans are 446.10\,s and 87.07\,s, respectively,
a 5.12-fold difference. The configurations send the same aggregate remote
payload but differ in route work and resulting makespan.
A widest path optimizes its bottleneck rate without accounting for the
additional contention created by its other hops.

\begin{table}[t]
\centering
\caption{Routing comparison: makespan (seconds) for widest-path (W) vs.\
minimum-hop (M) under HEFT with \texttt{csma\_bianchi}.}
\label{tab:routing}
\begin{tabular}{@{}llrrr@{}}
\toprule
\textbf{Network} & \textbf{DAG} & \textbf{W (s)} & \textbf{M (s)} & \textbf{Winner} \\
\midrule
$2\times2$ & 5 & 14.87 & 14.87 & Tie \\
$2\times2$ & 10 & 28.49 & 28.49 & Tie \\
$2\times2$ & 20 & 57.36 & 49.83 & Min-hop \\
$3\times3$ & 5 & 8.33 & 8.33 & Tie \\
$3\times3$ & 10 & 27.39 & 27.39 & Tie \\
$3\times3$ & 20 & 112.88 & 78.73 & Min-hop \\
$4\times4$ & 5 & 108.04 & 21.24 & Min-hop \\
$4\times4$ & 10 & 179.43 & 34.77 & Min-hop \\
$4\times4$ & 20 & 446.10 & 87.07 & Min-hop \\
\bottomrule
\end{tabular}
\end{table}

\textbf{Key findings.}
Minimum-hop routing is faster in five of nine configurations, with four
ties. On the 4$\times$4 grid, widest-path makespans are 108.04, 179.43,
and 446.10\,s for 5, 10, and 20 tasks; minimum-hop gives 21.24, 34.77,
and 87.07\,s. On the smaller grids the differences are more limited.
The effect of routing on wireless load depends on task placement.

\subsection{Interference Impact Analysis}
\label{sec:interference_eval}

Before asking whether interference changes scheduler \emph{rankings},
we first quantify its effect on absolute makespan. We compare Full with
Solo for the nine grid/DAG pairs under minimum-hop routing and report HEFT
and CPOP separately in \Cref{tab:interference}. The same placement is used
in each mode. Across these eighteen policy comparisons, slowdown ranges
from zero to 298\%. The zero cases are informative: with no overlapping
remote demands, changing the concurrent-link model need not change service.

\begin{table}[t]
\centering
\caption{Interference impact: makespan (seconds) without interference vs.\
with Full wireless service, using minimum-hop routing.  HEFT and CPOP
produce identical placements on the 5-task DAG but diverge as DAG
complexity grows.}
\label{tab:interference}
\setlength{\tabcolsep}{2.5pt}
\begin{tabular}{@{}ll|rrr|rrr@{}}
\toprule
& & \multicolumn{3}{c|}{\textbf{HEFT}} & \multicolumn{3}{c}{\textbf{CPOP}} \\
\textbf{Net.} & \textbf{DAG} &
  \textbf{Solo} & \textbf{Full} & \textbf{Slow.} &
  \textbf{Solo} & \textbf{Full} & \textbf{Slow.} \\
\midrule
$2\times2$ & 5T & 14.87 & 14.87 & 0\% & 14.87 & 14.87 & 0\% \\
$2\times2$ & 10T & 25.26 & 28.49 & 13\% & 23.91 & 27.54 & 15\% \\
$2\times2$ & 20T & 31.22 & 49.83 & 60\% & 30.00 & 53.78 & 79\% \\
$3\times3$ & 5T & 8.33 & 8.33 & 0\% & 8.33 & 8.33 & 0\% \\
$3\times3$ & 10T & 15.72 & 27.39 & 74\% & 14.69 & 27.17 & 85\% \\
$3\times3$ & 20T & 24.44 & 78.73 & 222\% & 22.84 & 62.46 & 173\% \\
$4\times4$ & 5T & 12.49 & 21.24 & 70\% & 12.49 & 21.24 & 70\% \\
$4\times4$ & 10T & 20.50 & 34.77 & 70\% & 20.18 & 34.77 & 72\% \\
$4\times4$ & 20T & 29.09 & 87.07 & 199\% & 23.75 & 94.62 & 298\% \\
\bottomrule
\end{tabular}
\end{table}

\textbf{Key findings.}
The 3$\times$3, five-task case is colocated on the 300-unit/s node:
both modes complete in 8.33\,s. At twenty tasks, the same grid sends
remote dependencies, and HEFT increases from 24.44 to 78.73\,s
(222\%), while CPOP increases from 22.84 to 62.46\,s (173\%).
On the 4$\times$4 grid with twenty tasks, HEFT increases from 29.09
to 87.07\,s and CPOP from 23.75 to 94.62\,s. The latter is the
largest minimum-hop slowdown in the table, 298\%.

These differences depend on placement and temporal overlap, not just the
number of network links. They show why absolute prediction error and
scheduler-selection error are different questions: substantial slowdown
can occur without a change in the best scheduler.

\subsection{Scheduler Comparison and Rank Inversions}
\label{sec:scheduler_eval}

The interference impact results above suggest a deeper question: does
ignoring interference merely produce inaccurate makespan predictions, or
does it lead to wrong \emph{choices} among schedulers?  To answer this, we
execute a full factorial comparison: 3 networks $\times$ 3 DAGs $\times$
3 schedulers (HEFT, CPOP, RoundRobin) $\times$ 2 routings $\times$ 2
interference models $= 108$ simulation runs.  All runs use 802.11ax at
5\,GHz, heterogeneous compute capacities, and seed~42.

For each of the 18 (network, DAG, routing) triples, let $T_p^S$ and $T_p^F$
be the completed makespans of policy $p$ in Solo and Full mode. The eligible
policy set is fixed to HEFT, CPOP, and RoundRobin; all 108 runs complete.
Policies within $\max(10^{-6}\,\mathrm{s},10^{-6}T_{\min})$ of the minimum
form a winning set. We count a \emph{rank inversion} only when the Solo
and Full winning sets are disjoint. This avoids treating a change of name
within a tie as a different outcome.

For regret, choose the first Solo minimizer in the fixed order HEFT, CPOP,
RoundRobin and denote it by $\widehat p$. Then
\begin{equation}
\rho=\frac{T_{\widehat p}^F}{\min_{p\in\{\mathrm{H,C,RR}\}}T_p^F}-1.
\label{eq:regret}
\end{equation}
In this experiment tied Solo winners also have equal Full makespans, so
the reported regret does not depend on choosing between those tied policies.

\Cref{tab:winner_matrix} reports each policy's Full makespan relative
to the best policy in its scenario. The summary statistics are:
\begin{itemize}[itemsep=1pt,leftmargin=*]
\item \textbf{Rank inversion rate: 33.3\%} (6 of 18 triples).
\item \textbf{Mean relative regret: 5.6\%} across all 18 triples;
  median regret is zero.
\item \textbf{Maximum relative regret: 52.3\%} ($4\times4$ grid,
  five tasks, widest-path): the Solo-selected HEFT placement takes
  108.04\,s in Full, versus 70.94\,s for RoundRobin.
\end{itemize}
These statistics summarize the eighteen scenarios in the fixed factorial.

\begin{table*}[t]
\centering
\caption{Makespan ratios under \texttt{csma\_bianchi} relative to the best
  scheduler for each (network, DAG, routing) triple.  A ratio of 1.00
  denotes the best among the evaluated policies; higher values show how much worse that
  scheduler performs.  \emph{Solo selection} indicates which scheduler an
  interference-free evaluation would select.  H/C denotes a tied Solo winning set; all grid runs complete.}
\label{tab:winner_matrix}
\setlength{\tabcolsep}{3.5pt}
\begin{tabular}{@{}ll|rrrl|rrrl@{}}
\toprule
& & \multicolumn{4}{c|}{\textbf{Widest-path}} &
  \multicolumn{4}{c}{\textbf{Minimum-hop}} \\
\textbf{Net.} & \textbf{DAG} &
  \textbf{H} & \textbf{C} & \textbf{RR} & \textbf{Solo$\downarrow$} &
  \textbf{H} & \textbf{C} & \textbf{RR} & \textbf{Solo$\downarrow$} \\
\midrule
$2\times2$ & 5T & 1.00 & 1.00 & 2.49 & H/C & 1.00 & 1.00 & 1.83 & H/C \\
$2\times2$ & 10T & 1.03 & 1.00 & 2.58 & C & 1.03 & 1.00 & 1.58 & C \\
$2\times2$ & 20T & 1.00 & 1.14 & 1.85 & H & 1.00 & 1.08 & 1.60 & C \\
$3\times3$ & 5T & 1.00 & 1.00 & 4.03 & H/C & 1.00 & 1.00 & 4.03 & H/C \\
$3\times3$ & 10T & 1.01 & 1.00 & 8.34 & C & 1.01 & 1.00 & 4.97 & C \\
$3\times3$ & 20T & 1.11 & 1.00 & 4.37 & H & 1.26 & 1.00 & 4.14 & C \\
$4\times4$ & 5T & 1.52 & 1.52 & 1.00 & H/C & 1.00 & 1.00 & 3.23 & H/C \\
$4\times4$ & 10T & 1.02 & 1.02 & 1.00 & C & 1.00 & 1.00 & 3.51 & C \\
$4\times4$ & 20T & 1.00 & 1.19 & 1.34 & C & 1.00 & 1.09 & 6.00 & C \\
\bottomrule
\end{tabular}
\end{table*}

\textbf{HEFT or CPOP wins under Solo service.}
One or both is a Solo minimizer in all 18 triples. Their use of compute
capacity and route-dependent communication costs offers an advantage over
cyclic placement in these inputs. At five tasks they tie; at larger sizes
their placements and compute queues can differ.

\textbf{RoundRobin can win under Full service.}
On the 4$\times$4 grid with widest-path routing, RoundRobin wins at five
and ten tasks, but HEFT wins at twenty tasks. Its benefit in this grid therefore depends on DAG size.
At five tasks, HEFT and CPOP take 12.49\,s in Solo versus RoundRobin's
25.58\,s. In Full, their 108.04\,s makespan exceeds RoundRobin's
70.94\,s. Thus the preferred policy changes even though the same placements
and underlying single-link goodputs are used in each mode.

\textbf{Routing modulates the effect.}
The five-task 4$\times$4 case has 52.3\% regret with widest paths and zero
regret with minimum-hop paths. At twenty tasks on that grid, choosing CPOP
from Solo gives 19.1\% regret with widest paths and 8.7\% with minimum-hop
paths. Short routes reduce some coupling but do not remove every inversion.

\begin{figure}[t]
\centering
\includegraphics[width=\columnwidth]{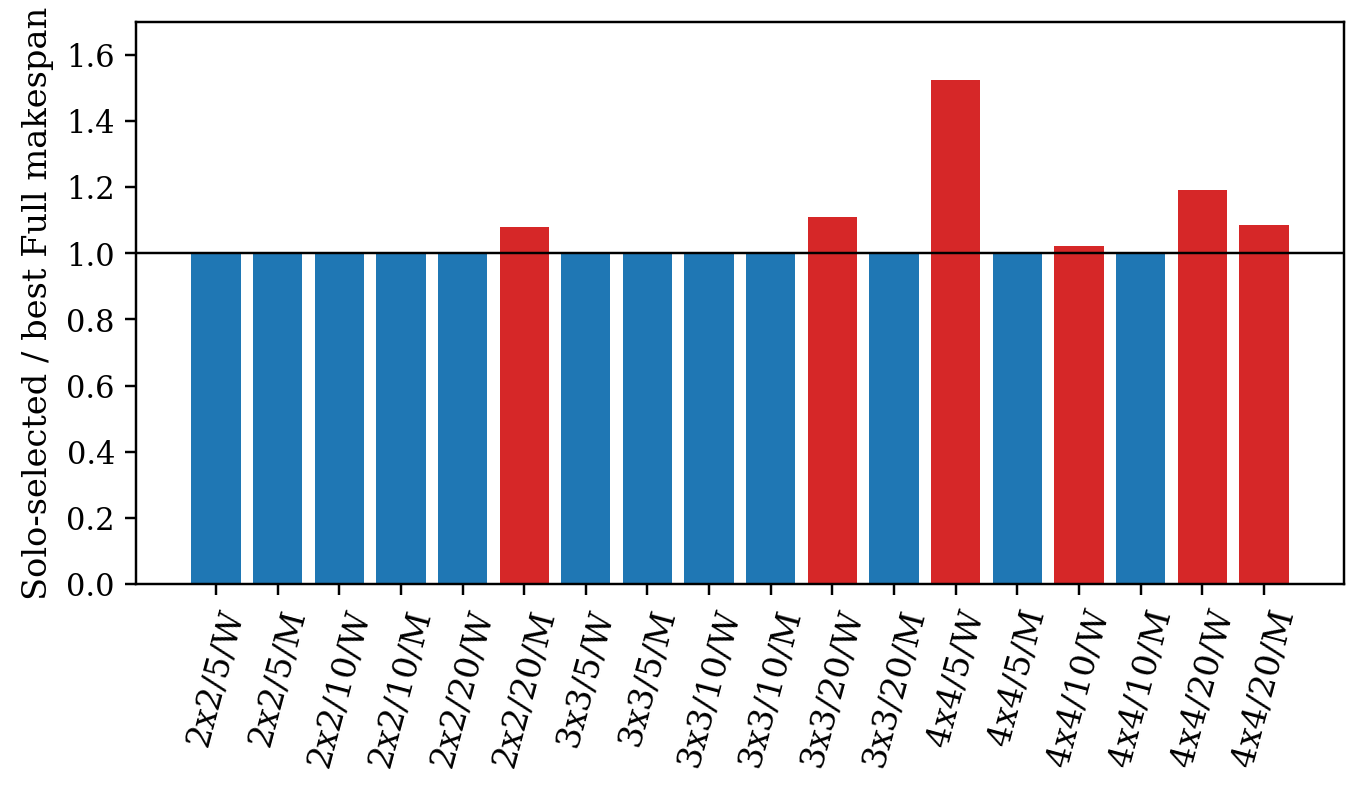}
\caption{Scheduling regret ratios. Each bar divides the Full makespan of
  the Solo-selected policy by the best evaluated-policy makespan under Full
  (HEFT, CPOP, or RoundRobin). A ratio of 1.0 indicates zero relative regret;
  red bars mark rank inversions. Labels give grid size/task count/routing,
  with W for widest-path and M for minimum-hop.}
\label{fig:regret_scatter}
\end{figure}

\Cref{fig:regret_scatter} shows all eighteen regret ratios, including the
twelve zero-regret cases. Six bars exceed one, and the largest is 1.523.
Each bar represents one grid, DAG, and routing combination.

In six of the eighteen tested scenarios, a policy selected under Solo
is no longer best among the evaluated policies under Full.

\subsection{Sensitivity Analysis}
\label{sec:sensitivity}

The results above show that interference can change scheduler rankings.
We vary data size per DAG edge to examine communication-to-computation
balance, and the number of injected DAGs to examine concurrent compute load.
The scheduler is rerun when payload size changes, so that sweep includes
both placement adaptation and runtime interference effects.

\subsubsection{Communication-to-Computation Ratio}

\begin{figure}[t]
\centering
\includegraphics[width=\columnwidth]{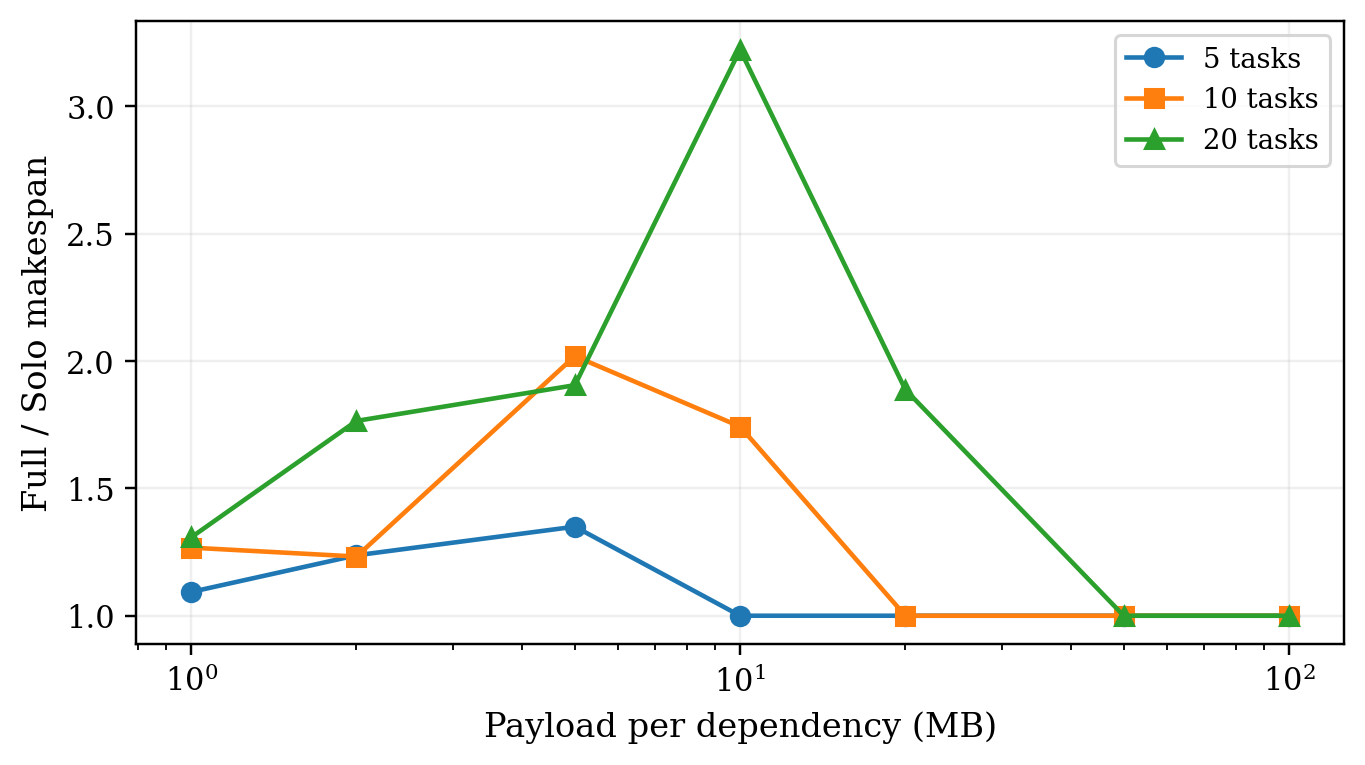}
\caption{Interference slowdown vs.\ payload per dependency (3$\times$3 grid,
  HEFT, minimum-hop). Slowdown peaks at intermediate payload sizes
  where transfers are long enough to contend but not so large that the
  scheduler collapses tasks onto one node.  The 20-task DAG reaches
  $3.22\times$ at 10\,MB per edge.}
\label{fig:ccr}
\end{figure}

The communication-to-computation ratio (CCR) determines the relative
importance of network transfer time versus compute time.  We sweep data
size per edge across $\{1, 2, 5, 10, 20, 50, 100\}$\,MB on the 3$\times$3
grid with HEFT and minimum-hop routing, testing all three DAG types
(5, 10, and 20 tasks).  \Cref{fig:ccr} shows the interference slowdown
as a function of payload size, the controlled communication-cost parameter.

The sweep shows a non-monotonic pattern. For the twenty-task
DAG, slowdown peaks at 3.22$\times$ with 10\,MB per edge: Solo makespan
is 24.44\,s and Full is 78.73\,s. At 50--100\,MB, HEFT colocates all
tasks on the fastest node, eliminates remote payload, and both modes take
33.33\,s. For five tasks this colocation occurs from 10\,MB, and for ten
tasks from 20\,MB.

At low payload sizes communication contributes less to completion time.
At high sizes the scheduler can avoid it through colocation. Intermediate
payloads can therefore create more exposure to wireless coupling. This is
a property of these freely placeable synthetic DAGs; memory or sensor-local
constraints could prevent complete colocation. We plot the controlled
payload size directly to identify the varied communication parameter.

\subsubsection{Concurrent Workflows and Compute Queuing}

As a compute-queuing control, we inject $k \in \{1, 2, 3, 4, 5\}$
identical 5-task fork-join DAGs with 0.5\,s stagger on the 3$\times$3 grid
(HEFT, minimum-hop).  Each DAG is scheduled independently, meaning that
HEFT optimizes each workflow without knowledge of the others' transfers.

\begin{figure}[t]
\centering
\includegraphics[width=\columnwidth]{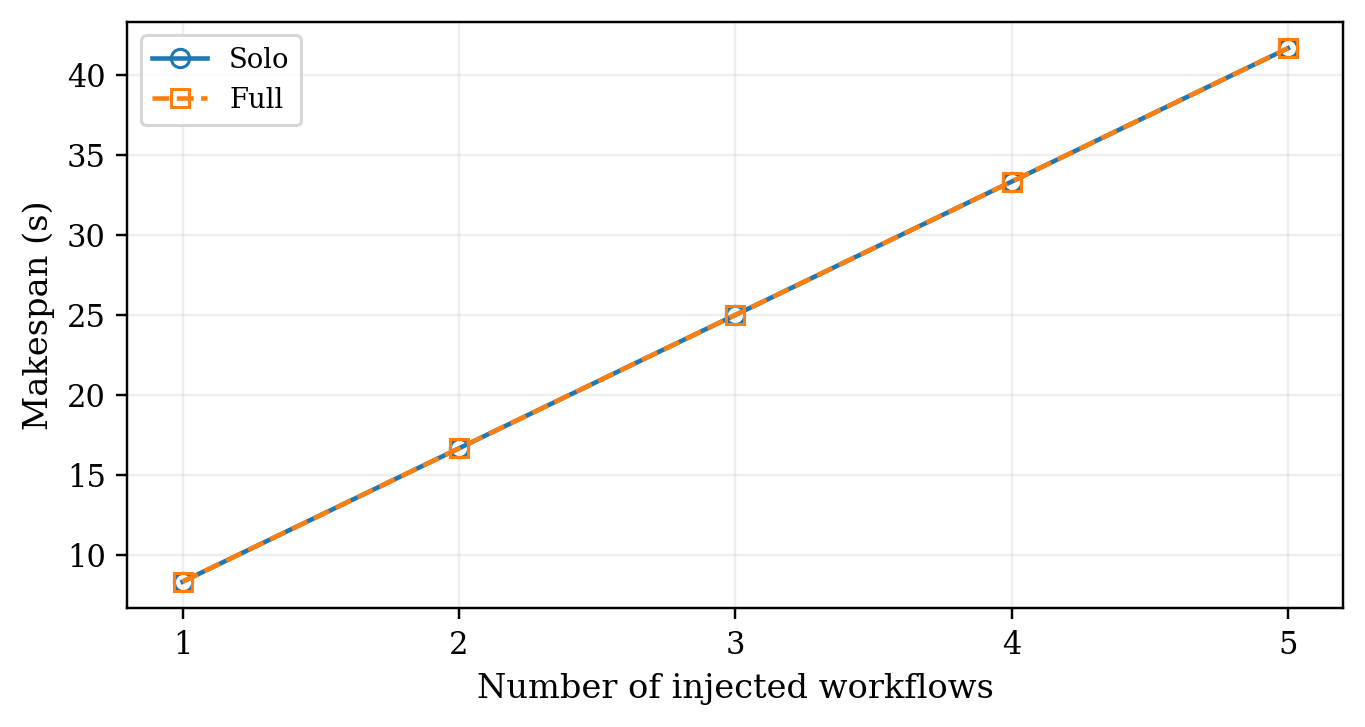}
\caption{Multi-DAG makespan scaling on the 3$\times$3 grid with
five-task DAGs, HEFT, and minimum-hop routing. Both modes colocate every
workflow on the fastest node and give $8.33k$\,s makespan. Coincident
curves show compute queuing rather than wireless contention.}

\label{fig:multidag}
\end{figure}

\Cref{fig:multidag} gives coincident Solo and Full curves. Each
five-task DAG is placed entirely on node $n_4$, the 300-unit/s node,
so no inter-node transfers are generated. One workflow requires
$5(500)/300=8.33$\,s of serial compute service. With 0.5\,s injection
spacing the processor remains busy, and five workflows require 41.67\,s.

This workload is compute-limited: colocated workflows share processor
capacity, so their makespan grows with the queued compute work.
Appendix~\ref{app:packet} separately uses prescribed starts and stops
of remote traffic to examine dynamic wireless response.

\subsubsection{Sensitivity Summary}

\Cref{tab:sensitivity_summary} consolidates the two sweeps. Moderate
payloads can expose a distributed placement to interference, while sufficiently
large payloads cause communication avoidance in these unconstrained DAGs.
The multi-DAG result shows how colocation shifts the shared load to compute.
For a given application, wireless load depends on where its tasks are placed
and which dependencies cross nodes.

\begin{table}[t]
\centering
\caption{Payload-size and workflow-count sensitivity.
The concurrent-workflow case generates no remote communication.}

\label{tab:sensitivity_summary}
\begin{tabular}{@{}llll@{}}
\toprule
\textbf{Experiment} & \textbf{Parameter} & \textbf{Observation} & \textbf{Effect} \\
\midrule
CCR & Edge MB & 20 tasks, 10 MB & $3.22\times$ slowdown \\
Multi-DAG & Count $k$ & all colocated & $8.33k$ s \\

\bottomrule
\end{tabular}
\end{table}

\section{Conclusion and Future Work}
\label{sec:conclusion}

\ncsim{} is a lightweight discrete-event simulator combining DAG workflow
scheduling with analytical 802.11 wireless service models. Validation
combines internal component checks, Bianchi's published table and figure,
and fixed-rate ns-3 experiments. The default model closely matches
homogeneous contention, and an optional fixed-capture mode reproduces the
hidden-terminal dip and recovery in the two-link fixed-MCS sweep.
Additional experiments examine overlapping contention domains and changing
traffic demand.

With a solo-MAC baseline, the 108-run
grid factorial yields six rank inversions out of eighteen scenarios
and a maximum scheduling regret of 52.3\%. The routing and CCR studies
show how communication demand interacts with placement.
The concurrent-workflow case colocates its tasks and gives
linear compute scaling.
For these freely placeable workloads, the routing and payload sweeps show
benefits from reducing routed communication or avoiding remote transfers
through colocation.

\textbf{Scope and limitations.} The model uses local homogeneous Bianchi
calculations, persistent hidden-interference power, fixed path loss, and
pipelined fluid transfers. It does not resolve packet retries, transport
feedback, mobility, or station-level transmit scheduling. Ready-time FIFO
execution also differs from enforcing a SAGA processor timeline.
The default model overestimates service in the tested hidden-interference
and overlapping-domain configurations (\Cref{sec:ns3_validation} and
Appendix~\ref{app:packet}). The workflow comparisons establish selection
sensitivity under this analytical service model; these experiments do not
establish packet-level agreement of scheduler rankings.

Future work will examine adaptive rate control and its interaction with
hidden interference. Extending the airtime model to account for contention
among hidden transmitters in larger multihop networks remains future work.
Other directions include receiver-specific service calibration, station-level
sharing, interference-aware placement under memory and sensor-local
constraints, and dynamic rescheduling. The project repository is
\url{https://github.com/ANRGUSC/ncsim}; artifact organization and reproduction
instructions are described in Appendix~\ref{app:reproduction}.

\section*{Acknowledgment}
This work was supported in part by Army Research Laboratory under Cooperative Agreement
W911NF-17-2-0196 and in part by the National Science Foundation under Award No. 2451267.

\section*{AI Use Acknowledgement}
We made extensive use of ChatGPT, Codex, Claude Code, Grok and Gemini for
software development, experiment execution, and manuscript preparation.
The human authors accept full responsibility for its contents.

\suppressfloats[t]
\appendices
\section{RF Parameters and Frame Timing}
\label{app:timing}

The experiments use 20\,dBm transmit power, 5\,GHz carrier frequency,
20\,MHz channel width, one spatial stream, path-loss exponent three,
$-95$\,dBm effective noise floor, and $-82$\,dBm CCA threshold.
Hidden-interference contributions below $-105$\,dBm are omitted. Static
shadowing is disabled, RTS/CTS is off unless stated, and scenario route
latencies are zero. The model's ax rate table is given in
\Cref{tab:mcs_parameters}. These thresholds and the 5\,dB effective-rate
margin are heuristic parameters. The margin acts as a single effective
offset to the rate-selection thresholds.

\begin{table}[t]
\centering
\caption{Single-stream ax clean-rate table at 20\,MHz. SNR thresholds are
model parameters; PHY rates are the rounded values used in execution.}
\label{tab:mcs_parameters}
\begin{tabular}{rrrrrr}
\toprule
MCS & SNR dB & Mb/s & MCS & SNR dB & Mb/s \\
\midrule
0 & 5 & 8.6 & 6 & 25 & 77.4 \\
1 & 8 & 17.2 & 7 & 29 & 86.0 \\
2 & 11 & 25.8 & 8 & 32 & 103.2 \\
3 & 14 & 34.4 & 9 & 35 & 114.7 \\
4 & 18 & 51.6 & 10 & 38 & 129.0 \\
5 & 22 & 68.8 & 11 & 41 & 143.4 \\
\bottomrule
\end{tabular}
\end{table}

The ax timing model uses a 1472-byte application payload and a 1542-byte
PSDU including MAC, LLC/SNAP, IP, UDP, FCS, and delimiter overhead.
For $R>0$ in Mb/s, data-frame duration in microseconds is
\begin{equation}
T_D(R)=44+13.6\left\lceil\frac{16+8(1542)+6}{13.6R}\right\rceil.
\end{equation}
The 44\,$\mu$s term is the preamble, and the 13.6\,$\mu$s symbol includes
an 800\,ns guard interval. The extra bits are service and tail bits.
For $W=16$, $m=6$, slot $\sigma=9$\,$\mu$s, SIFS $=16$\,$\mu$s,
AIFS $=43$\,$\mu$s, ACK duration $=28$\,$\mu$s, and propagation
$\delta=0.1$\,$\mu$s, the implemented basic-access durations are
\begin{align}
T_S&=T_D+\mathrm{SIFS}+T_{ACK}+\mathrm{AIFS}+2\delta,\\
T_C&=T_D+\mathrm{SIFS}+T_{ACK}+\mathrm{AIFS}+\delta.
\end{align}
The collision duration includes an extended inter-frame interval in this
approximation. Slot-model propagation is distinct from configured route
latency. ACK duration includes legacy OFDM control framing at 24\,Mb/s.
With RTS/CTS enabled, $T_{RTS}=T_{CTS}=28$\,$\mu$s and
\begin{align}
T_S^{RTS}&=T_S+T_{RTS}+T_{CTS}+2\mathrm{SIFS}+2\delta,\\
T_C^{RTS}&=T_{RTS}+\mathrm{SIFS}+T_{CTS}+\mathrm{AIFS}+\delta.
\end{align}
Thus rate changes alter efficiency as well as raw capacity. For example,
$\eta(1,8.6)=0.834788$ and $\eta(1,143.4)=0.279415$. The Bianchi
literature comparison instead uses the published FHSS parameters in
\Cref{sec:external_val}; the two timing settings are not interchangeable.

The fixed-point regression checks every contender count from 1 to 1000
against the coupled-equation residual. A separate 60-digit decimal solver
using the rational expression in the attempt-probability domain checks
selected counts, including 8, 20, 50, 100, 101, 256, and 1000.
These checks cover the solver used by both runtime and figure generation.

\subsection{Meaning of local conflict shares}

For an undirected active conflict graph with degree $d_\ell$, the shares
$1/(1+d_\ell)$ are graph-feasible: randomly order vertices and select
each vertex preceding all its neighbors. The selected set is independent,
and each vertex is selected with probability $1/(1+d_\ell)$.
Time-sharing these sets, and omitting service where necessary, realizes
any smaller nonnegative vector. For positive effective rate, the implemented
payload service normalized by clean PHY rate satisfies
\begin{equation}
\frac{R_\ell^{\mathrm{eff}}}{R_\ell}
\frac{\eta(1+d_\ell,R_\ell^{\mathrm{eff}})}{1+d_\ell}
\leq\frac{1}{1+d_\ell}.
\end{equation}
At zero effective rate, the normalized service is zero by definition.
This establishes feasibility under the abstract conflict constraints.
DCF service allocation is assessed separately through the overlapping-domain
packet comparisons in Appendix~\ref{app:packet}.

\section{Event-Engine Examples and Checks}
\label{app:engine}

A small example illustrates the pipelined route and FIFO semantics. Nodes
$a,b,c$ each provide one compute unit/s; links $a\to b$ and $b\to c$
each provide 2\,MB/s with zero latency and no wireless degradation.
Task A on $a$ requires one unit and produces 4\,MB for C on $c$.
Task B on $b$ requires two units and produces 1\,MB for D on $c$.
C and D each require one unit. A and B are ready at zero.

\begin{table}[t]
\centering
\caption{Worked two-hop transfer and FIFO execution. Both route links of
A's transfer are occupied throughout its byte-service phase.}
\label{tab:worked_engine}
\begin{tabular}{lp{5.4cm}}
\toprule
Time (s) & State and service \\
\midrule
0--1 & A and B compute on distinct nodes. \\
1--2 & A's 4\,MB transfer uses $a\to b\to c$ at 2\,MB/s; 2\,MB remains at 2\,s. \\
2--3 & B's 1\,MB transfer shares $b\to c$. Each flow gets 1\,MB/s. B's transfer finishes at 3\,s. \\
3--3.5 & A's last 1\,MB receives 2\,MB/s; D computes on $c$ from 3 to 4\,s. \\
3.5--5 & C becomes ready at 3.5\,s, waits for D, then runs from 4 to 5\,s. \\
\bottomrule
\end{tabular}
\end{table}

The workflow makespan is 5\,s. Link $a\to b$ serves 4\,MB and $b\to c$
serves 5\,MB. These values follow from crediting progress before replacing
a rate, and reserving the FIFO head before processing its start event.
A same-node dependency requires no transfer and only waits for compute
availability. A transfer's configured route latency, if positive, is paid
once before byte service and is not charged again after a rate update.

The update sequence is:
\begin{enumerate}[itemsep=1pt,leftmargin=*]
\item Apply a transfer start or finish to route occupancy.
\item Collect changed links and active links whose carrier-sense or hidden
      dependency neighborhoods contain a changed link.
\item Collect distinct affected flows and credit elapsed service at their
      stored old rates using \Cref{eq:progress}.
\item Evaluate all links on their routes and assign the new bottlenecks.
\item Invalidate old completions; schedule zero remaining work immediately,
      positive-rate work at its new finish time, and leave zero-rate work paused.
\end{enumerate}
The invariant is that all work before the event time is credited at the
old rate before any replacement completion is scheduled.

Tests include fixed-rate chains, fork-join graphs, shared links, temporary
blockage followed by recovery, mutually stalled transfers, and nonzero
route latency. A global-refresh oracle recomputes every link after each
event and agrees with local refresh in multihop/shared/hidden geometries
at 20, 80, and 120\,m separations, in both Solo and Full mode.
The comparison checks task times, rates, remaining bytes, and served
byte-hops, not only final makespan. Threshold tests verify isolated-link
invariance on both sides of all twelve ax thresholds, with margins
0, 3, 5, and 8\,dB and RTS/CTS off and on.

\section{Additional Packet-Level Comparisons}
\label{app:packet}

These experiments examine overlapping contention domains, rate-dependent
overhead, and changes in traffic demand.
All use ns-3.41 and twenty seeds per setting.
Error percentiles below describe sample goodput errors within each setting.

\subsection{Overlapping contention domains}

Three parallel 30\,m links at vertical positions 0, 60, and 120\,m give
overlapping carrier-sense neighborhoods: B conflicts with A and C, while
A and C are hidden relative to one another. This packet experiment forces
MCS~5 and measures traffic from 0.5 to 5\,s.
\Cref{tab:packet_asymmetric} shows lower service for the middle link in
both models. The flow model predicts
1.275\,MB/s for B versus a packet median of 0.520\,MB/s, identifying a
service-allocation difference in overlapping contention domains.

\begin{table}[t]
\centering
\caption{Overlapping domains: per-link goodput across twenty seeds.
IQR is the interquartile range; p95 is the sample 95th percentile of
absolute relative error with packet goodput as denominator.}
\label{tab:packet_asymmetric}
\small
\begin{tabular}{lrrr}
\toprule
Link & ncsim MB/s & ns-3 median [IQR] & Error p95 \\
\midrule
A & 1.942 & 3.17 [3.16, 3.19] & 39.5\% \\
B & 1.275 & 0.52 [0.50, 0.53] & 166.0\% \\
C & 1.942 & 3.18 [3.16, 3.19] & 39.3\% \\
\bottomrule
\end{tabular}

\end{table}

\subsection{Rate-dependent overhead}

Four isolated-link settings force MCS~0 or MCS~11 with RTS/CTS off or on.
The nodes are 1\,m apart, providing high SNR for either forced rate, and
the control mode is legacy 24\,Mb/s. Each run measures five seconds of
1472-byte UDP payload traffic. \Cref{tab:packet_rates} checks the overhead
calculation at two widely separated rates under high-SNR conditions.
The four per-setting p95 errors are 0.6--1.6\%.

\begin{table}[t]
\centering
\caption{Isolated-link overhead checks, with twenty packet seeds per row.}
\label{tab:packet_rates}
\small
\begin{tabular}{rlrrr}
\toprule
MCS & RTS & ncsim MB/s & ns-3 median [IQR] & Error p95 \\
\midrule
0 & off & 0.897 & 0.88 [0.88, 0.89] & 1.6\% \\
0 & on & 0.852 & 0.84 [0.84, 0.84] & 1.5\% \\
11 & off & 5.009 & 4.99 [4.98, 4.99] & 0.6\% \\
11 & on & 3.852 & 3.84 [3.84, 3.84] & 0.6\% \\
\bottomrule
\end{tabular}

\end{table}

\subsection{Controlled traffic starts and stops}

Two ad hoc 30\,m links are separated by 40\,m (mutual sensing) or 80\,m
(hidden interference). A offers 200\,Mb/s of 1472-byte UDP payloads from
0.5 to 5.5\,s, while B offers the same demand from 2 to 4\,s. Both force
MCS~5, use 24\,Mb/s control frames, and disable aggregation and RTS/CTS.
Static ARP entries avoid address-resolution startup. Network queue
disciplines are removed and each MAC queue is limited to one packet;
short and long retry limits are seven and four. These settings bound
drainage after source stop rather than assuming an instantaneous empty queue.

Received payload is recorded in 0.1\,s bins over 0.5--5.5\,s, including
transitions and zeros. The before/during/after windows in
\Cref{tab:dynamic_packet} are specified as 1--1.5, 2.5--3.5, and
4.5--5.5\,s. B delivers no payload in the final window in any seed.
Bins remain part of their seed's time series rather than independent samples.

\begin{figure}[t]
\centering
\includegraphics[width=\columnwidth]{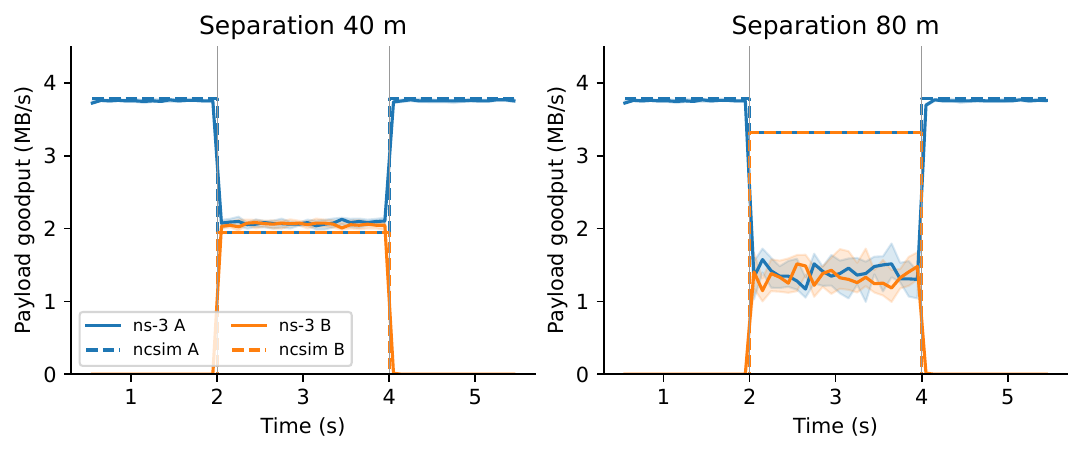}
\caption{Controlled UDP demand changes. Lines and bands show packet medians
and IQRs; dashed lines show active-set predictions. Vertical lines mark B's
source start and stop. The two geometries separate sensing from hidden interference.}
\label{fig:dynamic_packet}
\end{figure}

\begin{table*}[t]
\centering
\caption{Prespecified traffic windows. The twenty seeds per geometry supply
within-setting variability; zero-demand windows are retained.}
\label{tab:dynamic_packet}
\begin{tabular}{rlrrr}
\toprule
Spacing m & Link & Window s & ncsim MB/s & ns-3 median [IQR] MB/s \\
\midrule
40 & A & 1--1.5 & 3.783 & 3.757 [3.750, 3.765] \\
40 & A & 2.5--3.5 & 1.942 & 2.073 [2.048, 2.083] \\
40 & A & 4.5--5.5 & 3.783 & 3.757 [3.751, 3.764] \\
40 & B & 1--1.5 & 0.000 & 0.000 [0.000, 0.000] \\
40 & B & 2.5--3.5 & 1.942 & 2.060 [2.047, 2.086] \\
40 & B & 4.5--5.5 & 0.000 & 0.000 [0.000, 0.000] \\
80 & A & 1--1.5 & 3.783 & 3.757 [3.750, 3.765] \\
80 & A & 2.5--3.5 & 3.319 & 1.354 [1.319, 1.415] \\
80 & A & 4.5--5.5 & 3.783 & 3.758 [3.753, 3.765] \\
80 & B & 1--1.5 & 0.000 & 0.000 [0.000, 0.000] \\
80 & B & 2.5--3.5 & 3.319 & 1.362 [1.313, 1.407] \\
80 & B & 4.5--5.5 & 0.000 & 0.000 [0.000, 0.000] \\
\bottomrule
\end{tabular}

\end{table*}

At 40\,m, the flow prediction for A is 3.783, 1.942, and 3.783\,MB/s
before, during, and after B's demand, compared with packet medians of
3.757, 2.073, and 3.757\,MB/s. At 80\,m the model predicts 3.319\,MB/s
during overlap versus packet medians of 1.354 and 1.362\,MB/s for A and B.
Both models show reduced service when the competing flow starts and
recovery after it drains. During hidden-link overlap, the default-model
goodput is overestimated by about 2.4 times in this fixed-MCS test.

The main section evaluates 24 configurations (eight contention,
sixteen fixed separation), 480 configuration--seed runs, and 1360 measured
link observations. The appendix adds one overlapping-domain geometry,
four overhead settings, and two dynamic
geometries. Eight additional contention settings use a 4.5\,s window;
their measurements are stored separately from the 29.5\,s
curve. Packet evidence is reported by configuration rather than collapsed
into a single simulator-wide accuracy percentage.

\section{Artifact Organization and Reproduction}
\label{app:reproduction}

The study folder is \path{artifacts/arxiv-2605.01094/} in the
\ncsim{} repository. Within it, \path{inputs/} contains frozen workload
and topology generators, \path{results/workflows.json} contains the workflow
observations, and \path{results/packet/} contains seed-level packet data.
The \path{ns3/} directory supplies packet-experiment sources and build
instructions. \path{MANIFEST.json} records content hashes, and the README
specifies the software environment and reproduction commands.

The reported workflow studies comprise 108 grid-factorial executions,
42 payload-sensitivity executions, and ten multi-DAG executions.
Source hashes, generator hashes, Python 3.12.10, seed 42,
and hash seed zero are recorded. Figures and table rows are generated
from these records, and the internal parallel-link evidence is recomputed
using a separate piecewise-fluid calculation and the DES engine.

The Bianchi digitization records twelve manually identified marker centers,
source image hash, and pixel-to-axis calibration. Centers are read directly
from the plotted markers, independently of the calculated curves.
The fixed-MCS separation measurements use traffic from 0.5 to 30\,s,
with goodput computed over the full 29.5\,s interval.

Packet
comparisons measure UDP payload goodput under the configurations specified
in \Cref{sec:ns3_validation} and Appendix~\ref{app:packet}.

From the study folder, the following commands check the artifact, generate
figures and tables, run the workflow studies, validate the optional
fixed-MCS mode, and compile the manuscript:
\begin{lstlisting}[language=bash]
python reproduce.py verify
python reproduce.py figures
python reproduce.py workflows
python reproduce.py hidden
python reproduce.py paper
\end{lstlisting}
Generated files are written to a separate output directory, leaving the
saved observations unchanged. Figure generation uses saved workflow and
packet records and recomputes the small internal checks; compiling the
paper uses the supplied figure files. Packet simulations are run separately
using the instructions in \path{ns3/}. The manifest identifies the default
and optional models separately.

\bibliographystyle{IEEEtran}

\begin{thebibliography}{99}

\bibitem{shi2016edge}
W.~Shi, J.~Cao, Q.~Zhang, Y.~Li, and L.~Xu,
``Edge Computing: Vision and Challenges,''
\emph{IEEE Internet of Things Journal},
vol.~3, no.~5, pp.~637--646, Oct. 2016.

\bibitem{ns3}
{ns-3 Project},
``ns-3: A Discrete-Event Network Simulator,''
\url{https://www.nsnam.org/}, accessed Mar. 2026.

\bibitem{omnetpp}
A.~Varga and R.~Hornig,
``An Overview of the OMNeT++ Simulation Environment,''
in \emph{Proc. ICST SIMUTools}, 2008.

\bibitem{simgrid}
H.~Casanova, A.~Giersch, A.~Legrand, M.~Quinson, and F.~Suter,
``Versatile, Scalable, and Accurate Simulation of Distributed Applications
and Platforms,''
\emph{Journal of Parallel and Distributed Computing},
vol.~74, no.~10, pp.~2899--2917, 2014, doi: 10.1016/j.jpdc.2014.06.008.

\bibitem{wrench}
H.~Casanova, S.~Pandey, J.~Oeth, R.~Tanaka, F.~Suter, and R.~Ferreira~da~Silva,
``WRENCH: A Framework for Simulating Workflow Management Systems,''
in \emph{Proc. WORKS Workshop}, 2018, pp.~74--85, doi: 10.1109/WORKS.2018.00013.

\bibitem{cloudsim}
R.~N.~Calheiros, R.~Ranjan, A.~Beloglazov, C.~A.~F.~De~Rose, and R.~Buyya,
``CloudSim: A Toolkit for Modeling and Simulation of Cloud Computing
Environments and Evaluation of Resource Provisioning Algorithms,''
\emph{Software: Practice and Experience},
vol.~41, no.~1, pp.~23--50, 2011, doi: 10.1002/spe.995.

\bibitem{ifogsim}
H.~Gupta, A.~Vahid~Dastjerdi, S.~K.~Ghosh, and R.~Buyya,
``iFogSim: A Toolkit for Modeling and Simulation of Resource Management
Techniques in the Internet of Things, Edge and Fog Computing Environments,''
\emph{Software: Practice and Experience},
vol.~47, no.~9, pp.~1275--1296, 2017, doi: 10.1002/spe.2509.

\bibitem{edgecloudsim}
C.~Sonmez, A.~Ozgovde, and C.~Ersoy,
``EdgeCloudSim: An Environment for Performance Evaluation of Edge Computing
Systems,''
\emph{Transactions on Emerging Telecommunications Technologies},
vol.~29, no.~11, e3493, 2018.

\bibitem{bianchi2000}
G.~Bianchi,
``Performance Analysis of the IEEE 802.11 Distributed Coordination Function,''
\emph{IEEE Journal on Selected Areas in Communications},
vol.~18, no.~3, pp.~535--547, Mar. 2000, doi: 10.1109/49.840210.

\bibitem{heft}
H.~Topcuoglu, S.~Hariri, and M.-Y.~Wu,
``Performance-Effective and Low-Complexity Task Scheduling for Heterogeneous
Computing,''
\emph{IEEE Trans.\ Parallel Distrib.\ Syst.},
vol.~13, no.~3, pp.~260--274, 2002.

\bibitem{Adam1974ListSchedules}
T.~L.~Adam, K.~M.~Chandy, and J.~R.~Dickson,
``A Comparison of List Schedules for Parallel Processing Systems,''
\emph{Communications of the ACM},
vol.~17, no.~12, pp.~685--690, 1974.

\bibitem{KwokAhmad1999Survey}
Y.-K.~Kwok and I.~Ahmad,
``Static Scheduling Algorithms for Allocating Directed Task Graphs to
Multiprocessors,''
\emph{ACM Computing Surveys},
vol.~31, no.~4, pp.~406--471, 1999.

\bibitem{Sih1993DLS}
G.~C.~Sih and E.~A.~Lee,
``A Compile-Time Scheduling Heuristic for Interconnection-Constrained
Heterogeneous Processor Architectures,''
\emph{IEEE Trans.\ Parallel Distrib.\ Syst.},
vol.~4, no.~2, pp.~175--187, 1993.

\bibitem{Coleman2024PISA}
J.~Coleman and B.~Krishnamachari,
``PISA: An Adversarial Approach to Comparing Task Graph Scheduling Algorithms,''
in \emph{Proc. IEEE IPDPS}, pp.~54--66, 2025,
doi: 10.1109/IPDPS64566.2025.00014.

\bibitem{SAGARepo}
J.~Coleman, B.~Krishnamachari, and contributors,
``{SAGA}: Scheduling Algorithms Gathered (software),''
\url{https://github.com/anrgusc/saga}, accessed Sept.~2026.

\bibitem{Coleman2024Parametric}
J.~Coleman, R.~V.~Agrawal, E.~Hirani, and B.~Krishnamachari,
``Parameterized Task Graph Scheduling Algorithm for Comparing Algorithmic
Components,''
arXiv preprint arXiv:2403.07112, 2024.

\bibitem{Ghosh2021Jupiter}
P.~Ghosh, Q.~Nguyen, P.~K.~Sakulkar, J.~A.~Tran, A.~Knezevic, J.~Wang,
Z.~Lin, B.~Krishnamachari, M.~Annavaram, and S.~Avestimehr,
``Jupiter: A Networked Computing Architecture,''
in \emph{Proc.\ UCC Companion}, 2021, doi: 10.1145/3492323.3495630.

\bibitem{Poylisher2021TacticalJupiter}
A.~Poylisher, A.~Cichocki, K.~Guo, J.~Hunziker, L.~Kant,
B.~Krishnamachari, S.~Avestimehr, and M.~Annavaram,
``Tactical Jupiter: Dynamic Scheduling of Dispersed Computations in
Tactical MANETs,''
in \emph{MILCOM 2021}, 2021, pp.~102--107, doi: 10.1109/MILCOM52596.2021.9652937.

\bibitem{Zhao2021TPHEFT}
X.~Zhao, D.~Hu, and B.~Krishnamachari,
``Design and Experimental Evaluation of Algorithms for Optimizing the
Throughput of Dispersed Computing,''
arXiv preprint arXiv:2112.13875, 2021.

\bibitem{Suryavansh2020IBOT}
S.~Suryavansh, C.~Bothra, K.~T.~Kim, M.~Chiang, C.~Peng, and S.~Bagchi,
``{I-BOT}: Interference-Based Orchestration of Tasks for Dynamic Unmanaged
Edge Computing,''
arXiv preprint arXiv:2011.05925, 2020.

\bibitem{Li2022IBDASH}
X.~Li, M.~Abdallah, S.~Suryavansh, M.~Chiang, K.~T.~Kim, and S.~Bagchi,
``DAG-based Task Orchestration for Edge Computing,''
in \emph{Proc.\ SRDS}, 2022, pp.~23--34.

\bibitem{Li2024MTEC}
X.~Li, M.~Abdallah, Y.-Y.~Lou, M.~Chiang, K.~T.~Kim, and S.~Bagchi,
``Dynamic {DAG}-Application Scheduling for Multi-Tier Edge Computing in
Heterogeneous Networks,''
arXiv preprint arXiv:2409.10839, 2024.

\bibitem{Yi2017LAVEA}
S.~Yi, Z.~Hao, Q.~Zhang, Q.~Zhang, W.~Shi, and Q.~Li,
``{LAVEA}: Latency-aware Video Analytics on Edge Computing Platform,''
in \emph{Proc.\ SEC}, 2017.

\bibitem{Zhang2019HeteroEdge}
W.~Zhang, S.~Li, L.~Liu, Z.~Jia, Y.~Zhang, and D.~Raychaudhuri,
``Hetero-Edge: Orchestration of Real-time Vision Applications on
Heterogeneous Edge Clouds,''
in \emph{IEEE INFOCOM}, 2019, pp.~1270--1278.

\bibitem{Lin2019Petrel}
L.~Lin, P.~Li, J.~Xiong, and M.~Lin,
``Distributed and Application-aware Task Scheduling in Edge-clouds,''
arXiv preprint arXiv:1902.04362, 2019.

\bibitem{Liang2021JointOffloading}
J.~Liang, K.~Li, C.~Liu, and K.~Li,
``Joint offloading and scheduling decisions for {DAG} applications in
mobile edge computing,''
\emph{Neurocomputing},
vol.~424, pp.~160--171, 2021.

\bibitem{Li2024EnergyConstrained}
K.~Li,
``Energy-Constrained {DAG} Scheduling on Edge and Cloud Servers with
Overlapped Communication and Computation,''
\emph{Journal of Grid Computing}, vol.~22, Art.~60, 2024,
doi: 10.1007/s10723-024-09775-1.

\bibitem{Long2025SecDS}
L.~Long, Z.~Liu, J.~Shen, and Y.~Jiang,
``{SecDS}: A security-aware {DAG} task scheduling strategy for edge
computing,''
\emph{Future Generation Computer Systems}, vol.~166, Art.~107627, 2025.

\bibitem{Chiang2016FogIoT}
M.~Chiang and T.~Zhang,
``Fog and {IoT}: An Overview of Research Opportunities,''
\emph{IEEE Internet of Things Journal},
vol.~3, no.~6, pp.~854--864, 2016.

\bibitem{Chiang2017Fog10Q}
M.~Chiang, S.~Ha, C.-L.~I, F.~Risso, and T.~Zhang,
``Clarifying Fog Computing and Networking: 10 Questions and Answers,''
\emph{IEEE Communications Magazine},
vol.~55, no.~4, pp.~18--20, 2017.

\bibitem{Kim2020CodedEdge}
K.~T.~Kim, C.~Joe-Wong, and M.~Chiang,
``Coded Edge Computing,''
in \emph{IEEE INFOCOM}, 2020, pp.~237--246.

\bibitem{daneshgaran2008}
F.~Daneshgaran, M.~Laddomada, F.~Mesiti, M.~Mondin, and M.~Zanolo,
``Saturation Throughput Analysis of {IEEE} 802.11 in the Presence of Non
Ideal Transmission Channel and Capture Effects,''
\emph{IEEE Trans.\ Communications},
vol.~56, no.~7, pp.~1178--1188, Jul. 2008, doi: 10.1109/TCOMM.2008.060397.

\bibitem{zorzi1994}
M.~Zorzi and R.~R.~Rao,
``Capture and Retransmission Control in Mobile Radio,''
\emph{IEEE J.\ Sel.\ Areas Commun.},
vol.~12, no.~8, pp.~1289--1298, Oct. 1994, doi: 10.1109/49.329345.

\bibitem{hadzivelkov2002}
Z.~Hadzi-Velkov and B.~Spasenovski,
``Capture Effect in {IEEE} 802.11 Basic Service Area Under Influence of
{R}ayleigh Fading and Near/Far Effect,''
in \emph{IEEE PIMRC}, 2002.

\bibitem{simgridmodels}
SimGrid Project, ``The SimGrid Models,'' documentation, accessed Sept.~2026.
\url{https://simgrid.org/doc/latest/Models.html}.

\bibitem{flowwifi2022}
C.~Courageux-Sudan, L.~Gu\'egan, A.-C.~Orgerie, and M.~Quinson,
``A Flow-Level Wi-Fi Model for Large Scale Network Simulation,''
in \emph{Proc. ACM MSWiM}, 2022, doi: 10.1145/3551659.3559022.

\bibitem{milcom2026}
M.~Gutierrez, J.~Coleman, and B.~Krishnamachari,
``Locality Beats Widest-Path: DAG Scheduling under Wireless Edge Interference,''
in \emph{Proc. IEEE MILCOM}, 2026, to appear.

\end{thebibliography}

\end{document}